\documentclass[]{jkas} % Option for manuscript

\usepackage{stackengine,scalerel}

\def\year{2023} % publication year
\def\volume{56} % journal volume
\def\issue{1} % journal issue
\def\beginpage{1} % first page of article
\def\received{---} % date paper was received by JKAS (ex. October 12, 2022)
\def\accepted{---} % date of acceptance (ex. December 20, 2022)
\def\published{---} % date of publication (ex. January ??, 2023)
\date{Received \received; Accepted \accepted; Published \published}

\newcommand\sgra{Sgr\,A$^{*}$\,}
\newcommand\meight{M\,87$^{*}$\,}
\newcommand\ehtim{{\tt eht-imaging}\,}
\newcommand\ngehtsim{{\tt ngehtsim}\,}
\newcommand\nxcorr{$\rho_{\rm NX}$\,}

\DeclareRobustCommand{\okina}{%
  \raisebox{\dimexpr\fontcharht\font`A-\height}{%
    \scalebox{0.8}{`}%
  }%
}
\def\hawaii{Hawai\okina{}i\,}

\def\apjl{ApJL}%           % Astrophysical Journal, Letters 
\def\apjs{ApJS}%           % Astrophysical Journal, Supplement 
\def\ao{Appl.~Opt.}%           % Applied Optics 
\def\aap{A\&A}%           % Astronomy and Astrophysics 
\title{%
Enhanced imaging of \meight: Simulations with the EHT and extended-KVN
}

\author[1,2,3,$\star$]{Ilje~Cho}{0000-0001-6083-7521}
\author[4,5]{Jongho Park}{0000-0001-6558-9053}
\author[1,6]{Do-Young Byun}{0000-0003-1157-4109}
\author[1]{Taehyun Jung}{0000-0001-7003-8643}
\author[7,8]{Lindy Blackburn}{0000-0002-9030-642X}
\author[9]{Freek Roelofs}{0000-0001-5461-3687}
\author[10]{Andrew Chael}{0000-0003-2966-6220}
\author[7,8]{Dominic~W. Pesce}{0000-0002-5278-9221}
\author[7,8]{Sheperd~S. Doeleman}{0000-0002-9031-0904}
\author[7,$^\dag$]{Sara Issaoun}{0000-0002-5297-921X}
\author[11,12]{Jae-Young Kim}{0000-0001-8229-7183}
\author[13]{Junhan Kim}{0000-0002-4274-9373}
\author[3]{Jos\'e L. G\'omez}{0000-0003-4190-7613}
\author[5]{Keiichi Asada}{0000-0001-6988-8763}
\author[1,6]{Bong~Won Sohn}{0000-0002-4148-8378}
\author[1,6]{Sang-Sung Lee}{0000-0002-6269-594X}
\author[1]{Jongsoo Kim}{0000-0002-1229-0426}
\author[14,15]{Sascha Trippe}{0000-0003-0465-1559}
\author[2]{Aeree Chung}{}

\affil[1]{Korea Astronomy and Space Science Institute, Daedeok-daero 776, Yuseong-gu, Daejeon 34055, Republic of Korea}
\affil[2]{Department of Astronomy, Yonsei University, Yonsei-ro 50, Seodaemun-gu, Seoul 03722, Republic of Korea}
\affil[3]{Instituto de Astrof\'{\i}sica de Andaluc\'{\i}a-CSIC, Glorieta de la Astronom\'{\i}a s/n, E-18008 Granada, Spain}
\affil[4]{School of Space Research, Kyung Hee University, 1732 Deogyeong-daero, Giheung-gu, Yongin-si, Gyeonggi-do 17104, Republic of Korea}
\affil[5]{Institute of Astronomy and Astrophysics, Academia Sinica, P.O. Box 23-141, Taipei 10617, Taiwan, R. O. C.}
\affil[6]{University of Science and Technology (UST), Gajeong-ro 217, Yuseong-gu, Daejeon 34113, Republic of Korea}
\affil[7]{Center for Astrophysics~|~Harvard \& Smithsonian, 60 Garden Street, Cambridge, MA 02138, USA}
\affil[8]{Black Hole Initiative, Harvard University, 20 Garden Street, Cambridge, MA 02138, USA}
\affil[9]{Department of Astrophysics, Institute for Mathematics, Astrophysics and Particle Physics (IMAPP), Radboud University, P.O. Box 9010, 6500 GL Nijmegen, The Netherlands}
\affil[10]{Princeton Gravity Initiative, Princeton University, Jadwin Hall, Princeton, NJ 08544, USA}
\affil[11]{Department of Physics, Ulsan National Institute of Science and Technology (UNIST), 50 UNIST-gil, Eonyang-eup, Ulju-gun, Ulsan 44919, Republic of Korea}
\affil[12]{Max Planck Institute for Radio Astronomy, Auf dem Huegel 69, D-53121 Bonn, Germany}
\affil[13]{Department of Physics, Korea Advanced Institute of Science and Technology (KAIST), 291 Daehak-ro, Yuseong-gu, Daejeon 34141, Republic of Korea}
\affil[14]{Department of Physics and Astronomy, Seoul National University, Gwanak-gu, Seoul 08826, Republic of Korea}
\affil[15]{SNU Astronomy Research Center (SNUARC), Seoul National University, Gwanak-gu, Seoul 08826, Republic of Korea}
\def\corrauthor{%
I.~Cho, \email{icho@kasi.re.kr} \\
$\dag$ NASA Hubble Fellowship Program, Einstein Fellow. 
}

\def\runningauthor{%
Ilje~Cho et al.
}

\def\runningtitle{%
Enhanced imaging of \meight: Simulations with the EHT and extended-KVN
}

\def\keywords{%
quasars: supermassive black holes, galaxies: jets, techniques: interferometric
}

\def\abstracttext{%
The Event Horizon Telescope (EHT) has successfully revealed the shadow of the supermassive black hole, \meight, with an unprecedented angular resolution of approximately 20\,$\mu$as at 230\,GHz. However, because of limited short baseline lengths, the EHT has been constrained in its ability to recover larger-scale jet structures. 
The extended Korean VLBI Network (eKVN) is committed to joining the EHT from 2024 that can improve short baseline coverage. 
This study evaluates the impact of the participation of eKVN in the EHT on the recovery of the \meight jet. 
Synthetic data, derived from a simulated \meight model, were observed using both the EHT and the combined EHT$+$eKVN arrays, followed by image reconstructions from both configurations.
The results indicate that the inclusion of eKVN significantly improves the recovery of jet structures by reducing residual noise. 
Furthermore, jackknife tests, in which one or two EHT telescopes were omitted $-$ simulating potential data loss due to poor weather $-$ demonstrate that eKVN effectively compensates for these missing telescopes, particularly in short baseline coverage. 
Multi-frequency synthesis imaging at 86$-$230\,GHz shows that the EHT$+$eKVN array enhances the recovered spectral index distribution compared to the EHT alone and improves image reconstruction at each frequency over single-frequency imaging. 
As the EHT continues to expand its array configuration and observing capabilities to probe black hole physics more in depth, the integration of eKVN into the EHT will significantly enhance the stability of observational results and improve image fidelity. This advancement will be particularly valuable for future regular monitoring observations, where consistent data quality is essential.
}

\begin{document}
\jkashead %% set title, authors, abstract, etc.

%%%%%%%%%%%%%%%%%%%%%%%%%%%%%%%%%%%%%%%%%%%%%%%%%%%%%%%%%%%%%%%%%%%%%
%%% BEGIN MAIN TEXT HERE %%%%%%%%%%%%%%%%%%%%%%%%%%%%%%%%%%%%%%%%%%%%
%%%%%%%%%%%%%%%%%%%%%%%%%%%%%%%%%%%%%%%%%%%%%%%%%%%%%%%%%%%%%%%%%%%%%

\section{Introduction} \label{sec:intro}

%%% FIGURE %%%%%%%%%%%%%%%%%%%%%%%%%%%%%%%%%%%%%%%%%%%%%%%%%%%%%%%%%%%%%%%%%%%
\begin{figure*}[t]
\centering 
\includegraphics[width=\linewidth]{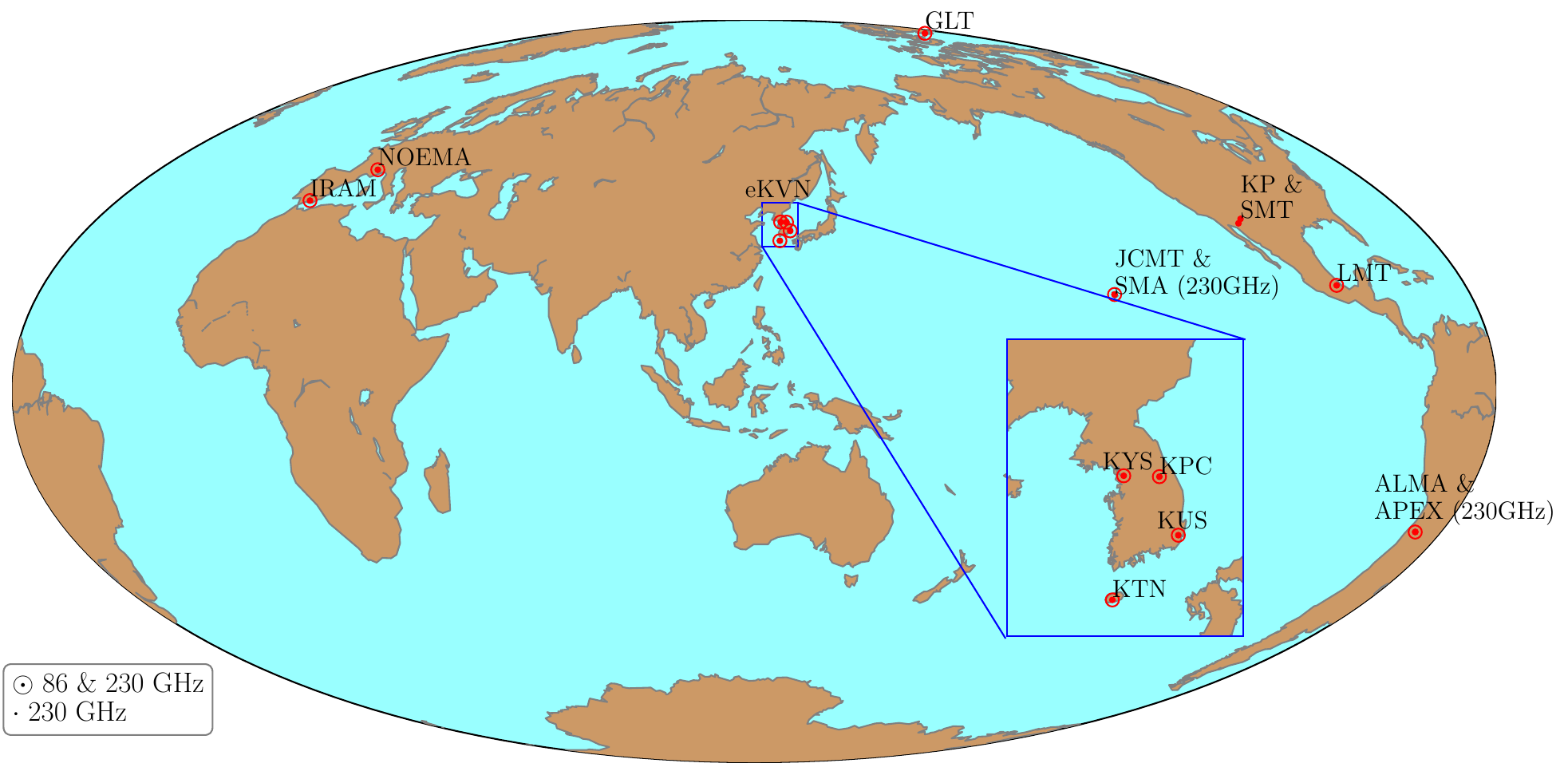}
\caption{
Locations of the EHT$+$eKVN telesctopes that are used for this study. The eKVN sites are zoomed in a inset box. 
}
\label{fig:globe}
\end{figure*}
%%% FIGURE %%%%%%%%%%%%%%%%%%%%%%%%%%%%%%%%%%%%%%%%%%%%%%%%%%%%%%%%%%%%%%%%%%%

%%%%%%%%%%%%%%%%%%%%%%%%%%%%%%%%%%%%%%%%%%%%%%%%%%%%%%%%%%%%%%%%%%%%%%%%%%%%%%%%%%%%%%%%%%%%%%
\begin{table*}[t]
\centering
\caption{EHT and eKVN telescopes toward \meight\,$^\S$}
\begin{tabular}{ccccccccc|cc}
\hline
\hline
Year & 2017 & 2018 & 2019 & 2020 & 2021 & 2022 & 2023 & 2024 & \multicolumn{2}{c}{2025$-$} \\
Frequency & \multicolumn{7}{c}{230\,GHz} && 86\,GHz (SEFD$^\dag$) & 230\,GHz (SEFD$^\dag$) \\
\hline
ALMA & O & O &  &  & O & O &  & O & O (69) & O (104) \\ 
APEX & O & O &  &  & O & O &  & O & -- & O (7727) \\ 
JCMT & O & O &  &  & O & O &  & O & O (3600) & O (5912) \\ 
SMA & O & O &  &  & O & O &  & O & -- & O (4954) \\ 
SMT & O & O &  &  & O & O &  & O & --$^{\dag\dag\dag}$ & O (8035) \\ 
IRAM & O & O &  &  & O & O &  & O & O (654) & O (1536) \\ 
LMT & O & O &  &  & -$^\ddag$ & O &  & O & O (513) & O (763) \\ 
GLT & - & O & -$^\ddag$ & -$^\ddag$ & O & O & (\sgra)$^{\dag\dag}$ & O & O (5312) & O (11231) \\ 
KP & - & - &  &  & O & O &  & O & --$^{\dag\dag\dag}$ & O (7823) \\ 
NOEMA & - & - &  &  & O & O &  & O & O (163) & O (398) \\ 
KYS & - & - &  &  & - & - &  & (\sgra)$^{\dag\dag}$ & O (3200) & O (35890) \\ 
KUS & - & - &  &  & - & - &  & - & O (3200) & $\bigtriangleup^{\dag\dag\dag}$ (35890) \\ 
KTN & - & - &  &  & - & - &  & - & O (3200) & $\bigtriangleup^{\dag\dag\dag}$ (35890) \\ 
KPC & - & - &  &  & - & - &  & - & O (3200) & O (35890) \\ 
\hline 
\end{tabular}
\vspace{0.3cm}
\raggedright{
\scriptsize{\\ 
$\S$ The participating telescopes are different depending on target source. For instance, the GLT is not available to observe \sgra due to its low declination but the South Pole Telescope (SPT) can join for \sgra.} \\
\scriptsize{
$\dag$ Expected zenith SEFDs in Jy, based on an optimal observing condition.} \\
\scriptsize{
$\ddag$ There were no observations in 2019 and 2020 due to COVID pandemic, and the LMT was missing in 2021 with the same reason.} \\
\scriptsize{
$\dag\dag$ Only \sgra sessions were carried out.} \\
\scriptsize{$\dag\dag\dag$ The 86\,GHz operations are planned at SMT and KP but not included yet in this study. The 230\,GHz receivers are planned to equip at KUS and KTN but it has been delayed compared to the other eKVN telescopes. In this study, all four eKVN telescopes are considered.} \\
}
\label{tab:obsplan}
\end{table*}
%%%%%%%%%%%%%%%%%%%%%%%%%%%%%%%%%%%%%%%%%%%%%%%%%%%%%%%%%%%%%%%%%%%%%%%%%%%%%%%%%%%%%%%%%%%%%%

% M87
The first-ever image of the black hole shadow was unveiled from the supermassive black hole (SMBH) at the heart of an elliptical galaxy, M\,87, through observations conducted with the Event Horizon Telescope (EHT; \citealt{ehtc_2019_1,ehtc_2019_2,ehtc_2019_3,ehtc_2019_4,ehtc_2019_5,ehtc_2019_6}). 
This landmark achievement validates the predictions from General Relativity under a strong gravitational field in the vicinity of black holes. 
At the same time, \meight is notable for its powerful jet extending even beyond its host galaxy. 
Especially with very long baseline interferometry (VLBI), the jet has been extensively studied to understand the mechanisms of astrophysical jet launching, collimation, and acceleration (see, e.g., \citealt{Asada_2012, Park_2019}). 
The jet launching mechanism is one of the longstanding questions in astrophysics, with theories proposing either a spinning black hole (BZ process; \citealt{blandford_1977}) or differential rotation of accretion flow (BP process; \citealt{blandford_1982}). 
Modeling of the observed black hole shadow based on General Relativistic Magnetohydrodynamic (GRMHD) simulations has suggested that the jet observed in \meight is likely launched by the BZ mechanism (e.g. \citealt{ehtc_2019_1}). However, the direct connection between the black hole shadow and the jet has yet to be confirmed through observation. 
Recently, at 86\,GHz, the ring-like shadow structure and larger scale jet were resolved simultaneously through observations conducted with the Global mm-VLBI Array (GMVA), in conjunction with the Atacama Large Millimeter/submillimeter Array (ALMA) (\citealt{lu_2023}). 
As a result, the edge-brightened width of the jet at the launching region exceeded the expected profile of the BZ process, suggesting the possibility of a wind originating from the accretion flow. 
It is therefore important to resolve the same structure at higher frequencies, examining the closer and deeper regions of the jet launching zone to further validate the jet launching mechanism. 
This can only be accomplished through the EHT with its future extension plan. 

% EHT
The EHT is a VLBI array operating at 230\,GHz with radio telescopes around the world (see, \autoref{fig:globe}), achieving an angular resolution of $\sim20\,\mu$as. 
Over the years, the number of participating telescopes that observe \meight has increased, from seven in 2017 to more than ten in later years (\autoref{tab:obsplan}), enhancing image fidelity by better covering the Fourier domain (that is, $uv-$coverage). 
In addition to expanding the array configuration, instrumental upgrades have also been implemented. 
For instance, in 2018, the total bandwidth doubled to 8\,GHz (except for the Greenland Telescope, GLT), and the effective dish diameter at the Large Millimeter Telescope Alforson Serrano (LMT) was increased by $\sim1.5$\,times compared to 2017 (\citealt{eht2018_M87_P1}). 
However, reconstructing the large-scale jet structure of \meight remains challenging.
This arises from the lack of short baseline lengths at $uv-$distance less than $\sim1\,$G$\lambda$, where $\lambda$ is the observing wavelength. 
The participation of Kitt Peak (KP) and the Northern Extended Millimeter Array (NOEMA) partially addresses this issue by providing baseline lengths of $\sim0.1\,{\rm G}\lambda$ (with the Sub-Millimeter Telescope, SMT) and $\sim0.8\,{\rm G}\lambda$ (with the IRAM 30\,m telescope), respectively, in later EHT observations. 
Nonetheless, there is still a gap of $\sim0.7\,{\rm G}\lambda$ in the short baseline coverage and a risk of losing the short baselines due to adverse weather conditions.

% KVN
The extended Korean VLBI Network\footnote{\url{https://radio.kasi.re.kr/status_report.php?cate=KVN}} (eKVN) comprises four 21\,m radio telescopes in South Korea (see \autoref{fig:globe} and \ref{app:ekvn_sites} for more details), an upgrade from the previous KVN configuration, with the addition of the KVN Pyeongchang (KPC) telescope and a 230\,GHz receiver\footnote{KVN Yonsei participated in the EHT campaign in April~2024 at 230\,GHz for the first time.}. 
This provides the shortest and longest baseline lengths of $\sim130\,$km (KVN Yonsei, KYS $-$ KPC) and $\sim500\,$km (KPC $-$ KVN Tamna, KTN), respectively. 
Through its participation in the EHT, the eKVN can contribute short baseline coverage of $\lesssim0.4\,{\rm G}\lambda$ at 230\,GHz corresponding to an angular scale of $\sim2.1-0.5$\,mas. 
This will be of great help recover the diffuse, large scale structure that has been challenging by far. 
Furthermore, eKVN is a pioneering VLBI array with a quasi-optics receiver system, enabling simultaneous observations at 22, 43, 86, 129\,GHz \citep[e.g.,][]{Kim_2018}, while compact triple-band receiver (CTR; \citealt{Han_2017}) at 22, 43, 86\,GHz has been equipped at KPC. 
Future eKVN will be progressively upgraded to have both CTR and a dual-band receiver at 129 and 230\,GHz so that it can simultaneously observe five frequencies in total. 
With this, the visibility phases at higher frequencies can be calibrated with more stable solution at lower frequencies, so called the Frequency Phase Transfer (FPT; e.g., \citealt{rioja_2011}) technique. 
In addition, the frequency dependent position shifts can be measured more accurately by applying the Source Frequency Phase Referencing (SFPR; \citealt{rioja_2011}) and the multi-frequency synthesis imaging can also be applied to obtain better images at each frequency, together with the spectral index distribution (see \autoref{subsec:mf_synthesis}). 
The EHT and the next generation EHT (ngEHT; \citealt{Doeleman_2023, Johnson_2023}) are also planning to equip the quasi-optics receiver covering from 86 to 345\,GHz \citep{Issaoun_2023, Rioja_2023, jiang_2023}. 
Therefore the EHT$+$eKVN observations can be implemented at both 86 and 230\,GHz simultaneously. 
Lastly, the ngEHT, including the eKVN, will further provide opportunities to investigate fundamental physics cases, such as the existence of the black hole photon ring ($n\geq1$), the first-ever spin measurement, tests of General Relativity, and the exploration of binary black holes \citep{Doeleman_2019, Ayzenberg_2023}. 

This paper aims to demonstrate the importance of eKVN in the future EHT observations through comparisons of reconstructed images of simulated \meight. 
In \autoref{sec:imsimul}, we introduce imaging simulations including groundtruth model, synthetic observational data, and image reconstruction procedures. 
Comparisons of reconstructed images at 230\,GHz and 86$+$230\,GHz multi-frequency synthesis are presented in \autoref{sec:results}. 
Based on the results, we discuss the implications of the improved images, the role of the eKVN, and future plans in \autoref{sec:summary}.

%%% FIGURE %%%%%%%%%%%%%%%%%%%%%%%%%%%%%%%%%%%%%%%%%%%%%%%%%%%%%%%%%%%%%%%%%%%
\begin{figure*}[t]
\centering 
\includegraphics[width=\linewidth]{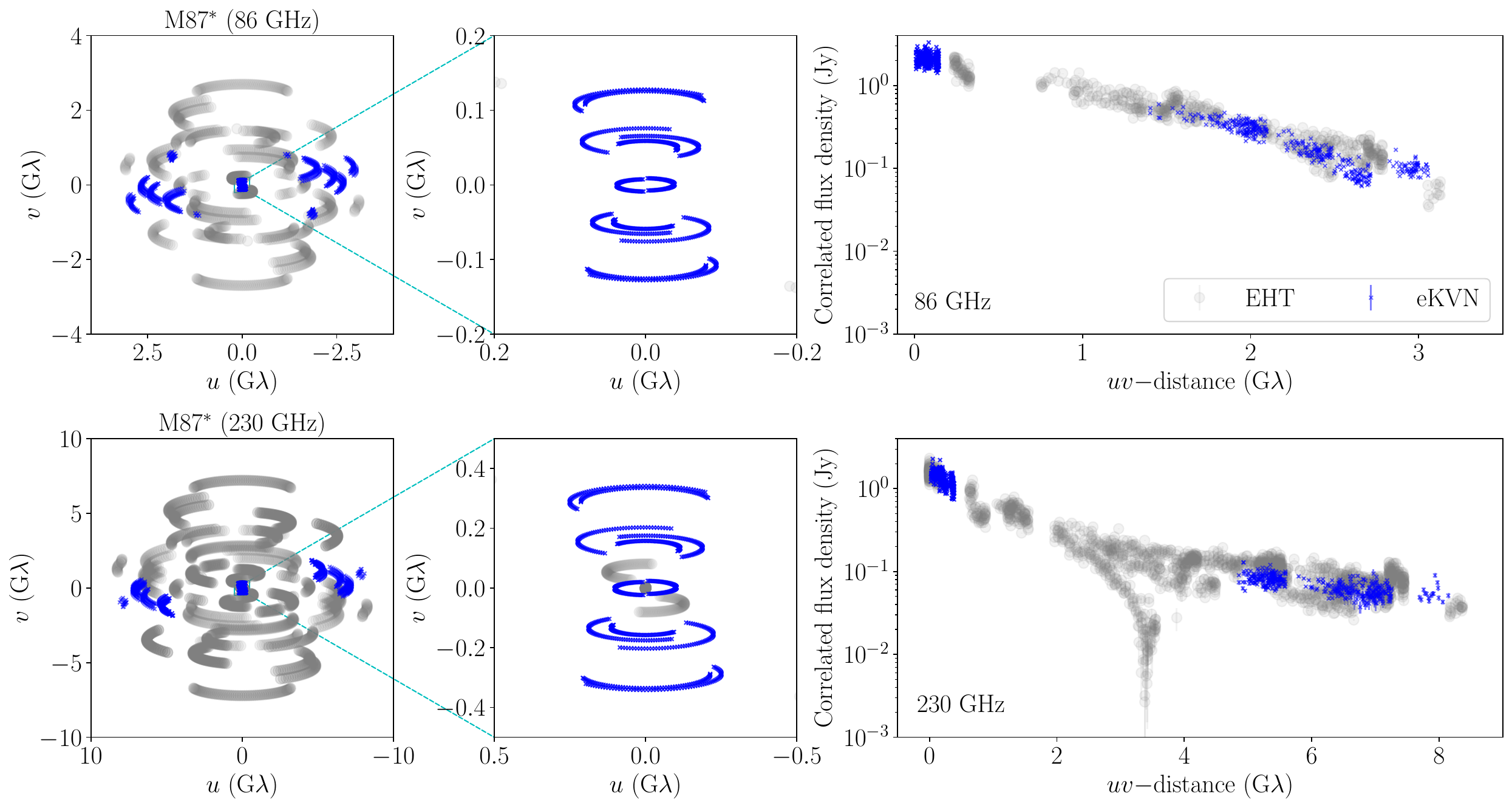}
\caption{
$uv-$coverage of full array (left) and short baselines region (middle) at 86\,GHz (top) and 230\,GHz (bottom). 
Gray and blue points are for EHT and eKVN baselines, respectively. 
The correlated flux density as a function of the $uv-$distance are shown together in right panel with same color-code. 
}
\label{fig:uvplt}
\end{figure*}
%%% FIGURE %%%%%%%%%%%%%%%%%%%%%%%%%%%%%%%%%%%%%%%%%%%%%%%%%%%%%%%%%%%%%%%%%%%

%%% FIGURE %%%%%%%%%%%%%%%%%%%%%%%%%%%%%%%%%%%%%%%%%%%%%%%%%%%%%%%%%%%%%%%%%%%
\begin{figure*}[ht!]
\centering 
\includegraphics[width=\linewidth]{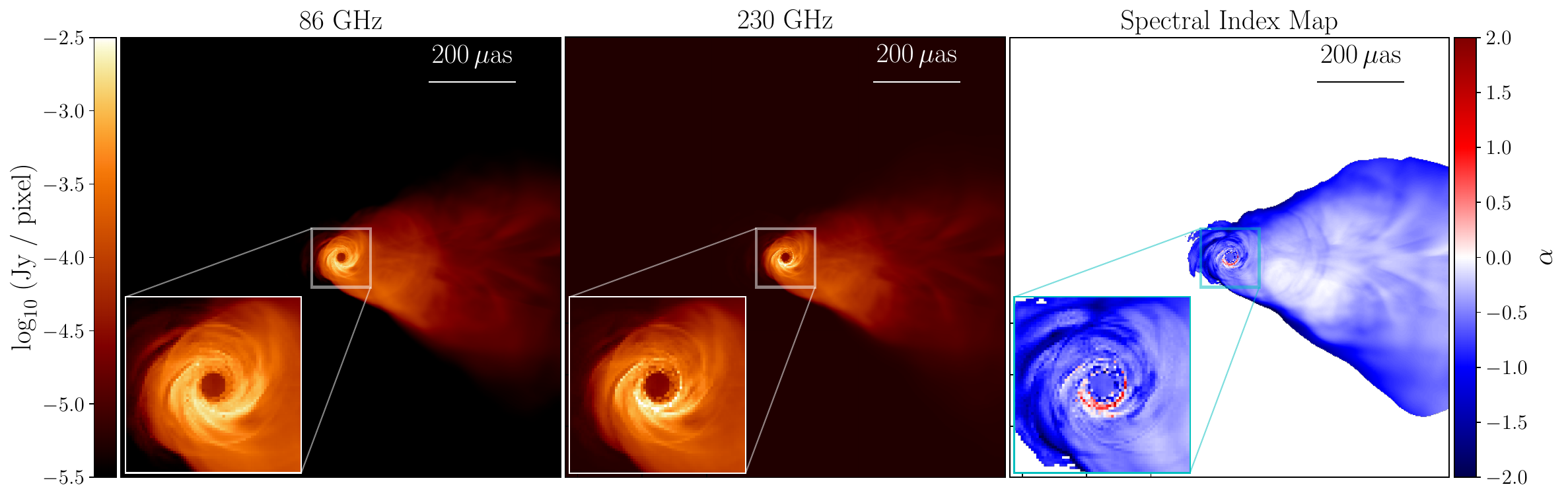}
\caption{
Simulated images of \meight (\citealt{Chael_2019}) at 86\,GHz (left), 230\,GHz (middle), and spectral index map between two frequencies (right). 
The central ring structure is zoomed in the inset at each panel. 
The images are not convolved with the observing angular resolution. 
}
\label{fig:groundtruth}
\end{figure*}
%%% FIGURE %%%%%%%%%%%%%%%%%%%%%%%%%%%%%%%%%%%%%%%%%%%%%%%%%%%%%%%%%%%%%%%%%%%

%%%%%%%%%%%%%%%%%%%%%%%%%%%%%%%%%%%%%%%%%%
\section{Imaging simulation} \label{sec:imsimul}

\subsection{Synthetic data generation} 
Synthetic data is generated using \ngehtsim software which considers realistic observing conditions based on historical atmospheric models and employs heuristic visibility detection criteria \citep{Pesce_2024, Pesce_2024zndo}. 
A simulated image of \meight is observed with the EHT, together with the eKVN, at 86 and 230\,GHz with 512\,MHz and 2\,GHz of bandwidths, respectively. 
Results from the partial participation of the eKVN (i.e., KYS and KPC) are also presented in \ref{app:ekvn_sites}. 
Note that the EHT telescopes are assumed to be in full array configuration as of 2022 (see, \autoref{tab:obsplan}), and a reference date for synthetic data generation is set for April~6, 2024. 
\autoref{fig:uvplt} presents the resultant $uv-$coverage (left, middle) and correlated flux density as a function of the $uv-$distance (right) of the synthetic data. 
The groundtruth model depicts both the black hole shadow and extended jet structure of \meight, based on GRMHD simulation (\citealt{Chael_2019}; \autoref{fig:groundtruth}). 
Full track observations for 24\,hours with an elevation limit between 10 to 80\,degrees are assumed, with a scan length of 5\,minutes and a gap of 10\,minutes between scans, mimicking real EHT observations (e.g., \citealt{ehtc_2019_3}). 
The fringe detection threshold is given by the signal-to-noise ratio (S/N) of 5, and the fringe coherence time is assumed as 10\,seconds (see, Appendix~G.1 in \citealt{Pesce_2024}, for more details). 
The gain corruptions at each station are estimated by the \ngehtsim, based on the expected weather condition (see, \ref{app:ekvn_sites}, for eKVN sites) and telescope's characteristics (e.g., gain curve).

\subsection{Imaging}
% Single frequency imaging
Images from the synthetic data were reconstructed using \ehtim which utilizes Python modules to manipulate and simulate VLBI data, and reconstruct images using regularized maximum likelihood (RML) methods (\citealt{Chael_2018}). 
First, data points with S/N lower than 3 were flagged out. 
Then image is found by minimizing the objective function,  
\begin{equation}
    J(I) = \underset{\rm data}\Sigma\kappa_{\rm D}\chi^{2}_{\rm D}(V,I) - \underset{\rm regularizers}\Sigma\lambda_{\rm R}S_{\rm R}(I),
\end{equation}
where $\chi^{2}_{\rm D}$ is likelihood that compares data products, $V$, and reconstructed image, $I$. $S_{\rm R}$ is regularizer, $\kappa_{\rm D}$ and $\lambda_{\rm R}$ are hyperparameters for each term. Data products include complex visibility, visibility amplitudes, closure phases, and closure amplitudes. Regularizer terms control the image features favoring such as sparsity and/or smoothness of edges that play as a penalty term for likelihood. 

In our imaging, a field-of-view was set as 1024\,$\mu$as with a pixel size of 4\,$\mu$as. 
For the initial model, a simple circular Gaussian brightness distribution was used. 
With this, the first image was obtained from data products of closure phases and logarithm of the closure amplitudes which were more robust as they were free from the gain corruptions. 
Both visibility amplitudes and phases were then self-calibrated with the images. 
Note that total flux of the images were not well constrained, as only closure quantities were used, so the overall fluxes were rescaled before self-calibration based on the groundtruth flux density \citep[e.g.,][]{Chael_2023}. 
After self-calibration, the visibility amplitudes were added to the data products for next imaging with a lower value of hyperparameter, $\kappa_{\rm D}$, than the closure quantities. 
This was repeated two times, together with the intervening self-calibration. 
To find the best combination of $\lambda_{\rm R}$, a parameter search was conducted \citep[e.g.,][]{ehtc_2019_4} in a range of [0.1, 1, 10] for each regularizer: (i) $l1$ norm which favors spatial sparsity, (ii) relative entropy (MEM) which favors similarity to a prior image\footnote{a circular Gaussian prior with a size of 50\,$\mu$as is used in this study.}, (iii) logarithm of total squared variation (TSV) introducing smooth edges, and (iv) source size regularizer ({\tt compact2}). 
As a result, 81 images were obtained and the fiducial image was selected based on the normalized cross correlation (\nxcorr) to the groundtruth image (see, e.g., \citealt{ehtc_2019_4}). 
Note that the groundtruth image was convolved with a circular Gaussian of 8\,$\mu$as diameter, corresponding to the effective angular resolution of the reconstructed images (see, \ref{app:beam}), and all reconstructed images were similar to each other providing the difference in \nxcorr of less than 10\%. 
Consequently, the fiducial parameters for the synthetic data generated by EHT-only observation were found to be $l1=0.1$, MEM$=$1, log(TSV)$=$0.1, and {\tt compact2}$=$1. 
To assess the effect of eKVN participation, avoiding differences caused by imaging parameters, the same fiducial parameters were applied to the synthetic data of the EHT$+$eKVN. 
The same set of parameters were also used for the jackknife test to assess the geographical effects of telescopes on each continent (see \autoref{sec:results}). 

% Multi frequency synthesis imaging
Regarding the multi-band receiver plan of the future EHT, multi-frequency synthesized imaging was also conducted at 86$+$230\,GHz. 
Note that the EHT will go up to 345\,GHz but our study does not consider it since this frequency is not suitable for the eKVN telescopes due to weather condition in South Korea (see, e.g., \ref{app:ekvn_sites}) and the aperture surface efficiency. 
The multi-frequency synthesis imaging is available with the \ehtim (\citealt{Chael_2023}), reconstructing images at both frequencies along with spectral index map, $\alpha$, and spectral curvature map, $\beta$, by minimizing the following objective function, 
\begin{equation}
    \begin{aligned}
    J(I_0,\alpha,\beta) & = \underset{\rm data}\Sigma 
    \underset{\,\,\nu_i}\Sigma \kappa_{\rm D}\chi^{2}_{\rm D}(V_i,I_i) \\
    & - \underset{\rm regularizers}\Sigma(\lambda_{\rm R}S_{\rm R}(I_0) + \lambda_{\alpha}S_{\alpha}{\rm (\alpha)} + \lambda_{\beta}S_{\beta}{\rm (\beta)}), 
    \end{aligned}
\end{equation}
where $V_{i}$ and $I_{i}$ represent data product and reconstructed image at different frequencies, $\nu_{i}$, respectively. 
The total intensity image regularizers are applied to the image at reference frequency, $\nu_0$, that is denoted as $I_0$. 
The spectral index and curvature maps are regularized by $S_{\alpha}{\rm (\alpha)}$ and $S_{\beta}{\rm (\beta)}$, respectively. 
The $\lambda_\alpha$ and $\lambda_\beta$ are corresponding hyperparameters. 

For multi-frequency synsthesis imaging of our synthetic data, 230\,GHz was selected as reference frequency since more telescopes are available than 86\,GHz. 
Note that, by this way, a better angular resolution can be obtained at 86\,GHz as well. 
The regularizers and corresponding hyperparameters of the reference frequency image, $\lambda_{\rm R}$, were adopted as a fiducial parameters combination from 230\,GHz single frequency imaging.  
For regularizing the spectral index map, $\alpha$, a parameter survey to find $\lambda_\alpha$ was conducted on two regularizers in a range of [1, 10, 30, 50]: 
(i) $l2$ norm that makes $\alpha$ converge on $\alpha_{0}$\footnote{$\alpha_{0}$ can be given as 0 or a measured value from total fluxes (i.e., unresolved source).} when there are no data constraints, and (ii) total variation (TV) preventing large variations in $\alpha$ over small parts of the image (see, \citealt{Chael_2023}, for more details). 
The spectral curvature map, $\beta$, was not found since only two frequencies were dealt with. 
After the survey, the fiducial parameters for $\alpha$ were selected as $l2=10$ and ${\rm TV}=50$ based on the highest \nxcorr at both frequencies, which were higher than the results from single frequency imaging. 

%%% FIGURE %%%%%%%%%%%%%%%%%%%%%%%%%%%%%%%%%%%%%%%%%%%%%%%%%%%%%%%%%%%%%%%%%%%
\begin{figure*}[t]
\includegraphics[width=\linewidth]{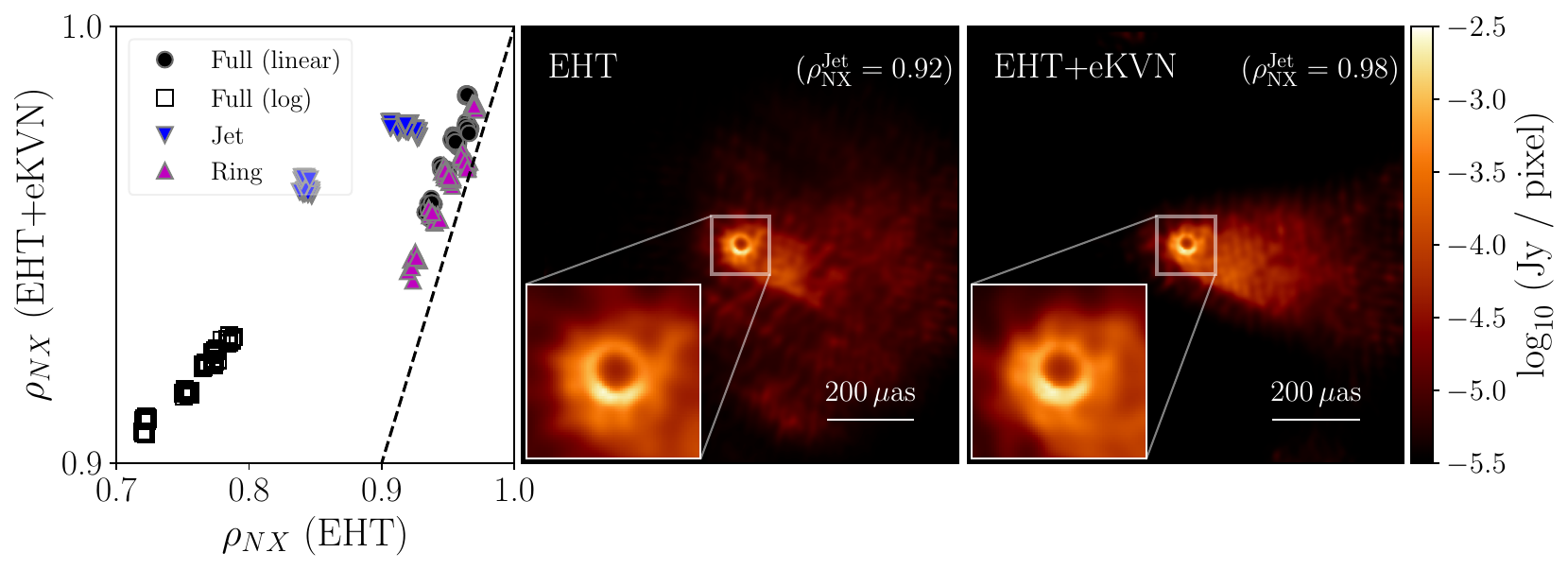} 
\vspace{-0.9cm}
\flushright\includegraphics[width=0.955\linewidth]{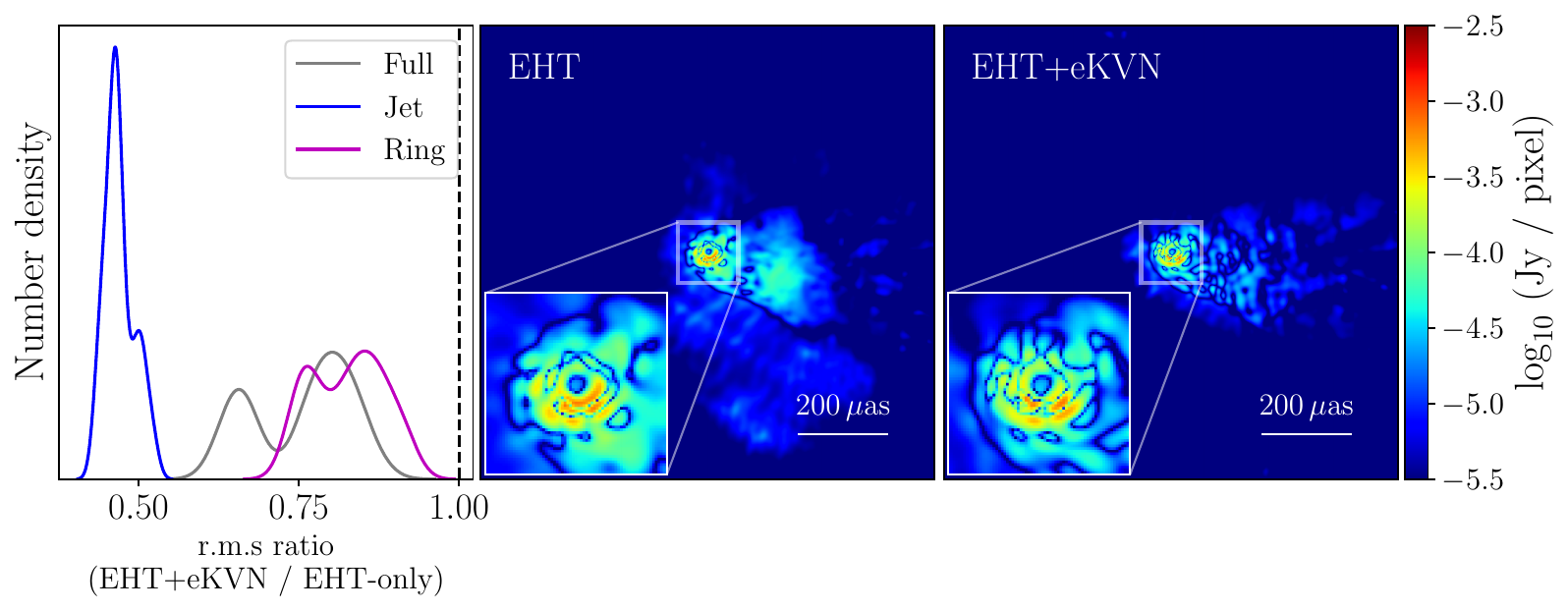}
\caption{
(Top) Imaging results at 230\,GHz. 
(Top, left) The \nxcorr comparison between EHT-only and EHT$+$eKVN reconstructions over the searched imaging parameters. 
Different colors and markers show the \nxcorr from full image in linear scale (black, circle), full image in log scale (black, square), ring (magenta, triangle), and jet (blue, down-pointing triangle) structures. 
Note that all \nxcorr from EHT$+$eKVN is higher than EHT-only cases. 
Especially the \nxcorr for jet region is improved more significantly by the eKVN participation, as demonstrated by $\rho_{\rm NX}^{\rm jet}$ and $\rho_{\rm NX,log}$.  
(Top, middle and right) Fiducial images at 230\,GHz from EHT-only and EHT$+$eKVN, respectively, in logarithmic scale. 
These are reconstructed by the same imaging parameters. 
(Bottom, left) The r.m.s ratio between images from EHT$+$eKVN and EHT-only. Note that the r.m.s implies difference between groundtruth and reconstructed images. The image differences are presented on middle and right panels for corresponding images in the top row. 
Note that the groundtruth image was blurred with a circular Gaussian of 8\,$\mu$as diameter before estimating the \nxcorr and r.m.s (see, \ref{app:beam}).} 
\label{fig:im_fiducial_nxcorr}
\end{figure*}
%%% FIGURE %%%%%%%%%%%%%%%%%%%%%%%%%%%%%%%%%%%%%%%%%%%%%%%%%%%%%%%%%%%%%%%%%%%

%%% FIGURE %%%%%%%%%%%%%%%%%%%%%%%%%%%%%%%%%%%%%%%%%%%%%%%%%%%%%%%%%%%%%%%%%%%
\begin{figure*}[ht]
\centering 
\includegraphics[width=\linewidth]{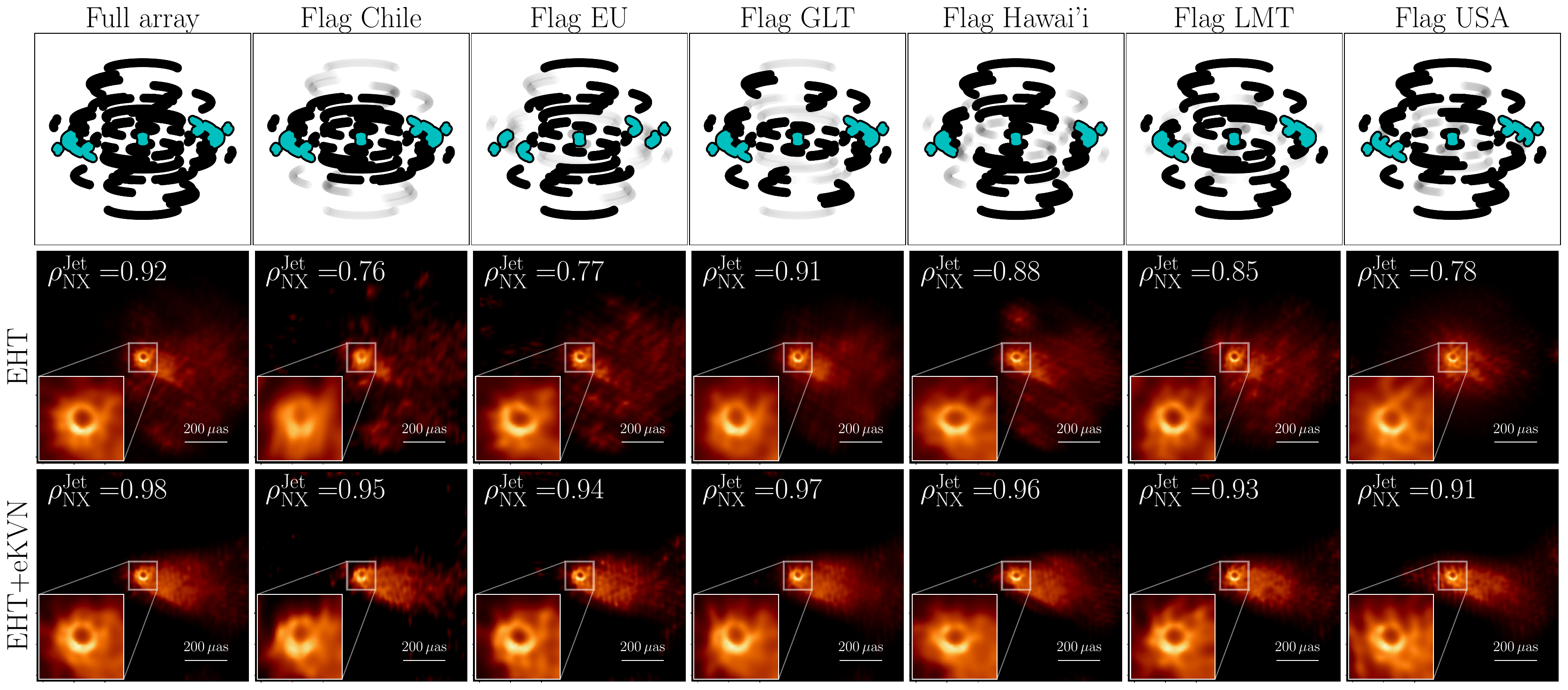}
\caption{
Jackknife test results of flagging different telescopes. 
From left to right, result from full array, flagging telescopes in Chile (ALMA, APEX), Europe (IRAM, NOEMA), Green land (GLT), \hawaii (JCMT, SMA), Mexico (LMT), and United States (KP, SMT). 
Top row presents the corresponding $uv-$coverage (black), together with the flagged points (faint gray) and the eKVN baselines (cyan). 
The second and third rows show the images from EHT-only and EHT$+$eKVN, respectively. 
The color scale is same with \autoref{fig:im_fiducial_nxcorr}. 
}
\label{fig:jackknife}
\end{figure*}
%%% FIGURE %%%%%%%%%%%%%%%%%%%%%%%%%%%%%%%%%%%%%%%%%%%%%%%%%%%%%%%%%%%%%%%%%%%

%%% FIGURE %%%%%%%%%%%%%%%%%%%%%%%%%%%%%%%%%%%%%%%%%%%%%%%%%%%%%%%%%%%%%%%%%%%
\begin{figure}[ht!]
\centering 
\includegraphics[width=0.95\linewidth]{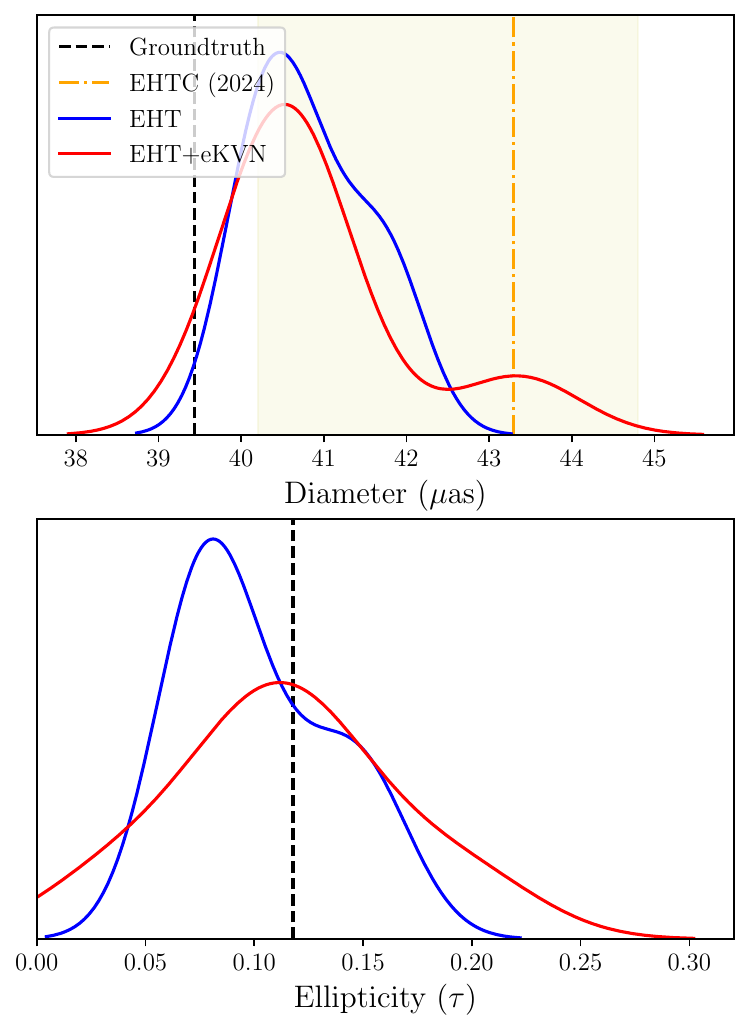}
\caption{
Ring fitting results on each of jackknife test, as shown in \autoref{fig:jackknife}, for diameter (top) and ellipticity (bottom). 
The EHT-only and EHT$+$eKVN results are shown with blue and red, respectively. 
Black-broken, vertical line presents the fitted value on groundtruth image. 
As for diameter, the result from \citealt{eht2018_M87_P1} onto real \meight is shown for comparison. 
}
\label{fig:idfe}
\end{figure}
%%% FIGURE %%%%%%%%%%%%%%%%%%%%%%%%%%%%%%%%%%%%%%%%%%%%%%%%%%%%%%%%%%%%%%%%%%%

%%%%%%%%%%%%%%%%%%%%%%%%%%%%%%%%%%%%%%%%%%
\section{Results} \label{sec:results}

\subsection{230\,GHz}
The normalized cross correlation, \nxcorr, of reconstructed images ranges from 0.92$-$0.99 (see \autoref{fig:im_fiducial_nxcorr}, top, left). 
For the same imaging parameters, the images from the EHT$+$eKVN exhibit higher \nxcorr compared to the EHT-only results. 
Since the central ring structure representing the black hole shadow is brighter than the downstream jet, the ring flux densities dominate the estimate for \nxcorr. 
To quantify the image reconstruction of fainter jet region, \nxcorr has been also estimated for the ring ($\rho_{\rm NX}^{\rm ring}$) and the jet ($\rho_{\rm NX}^{\rm jet}$) region separately by masking out and in a circle with a 70\,$\mu$as diameter at the center, respectively. 
As a result, the \nxcorr of both structures, as well as the entire image, are higher in EHT$+$eKVN results. 
Especially the improvement in $\rho_{\rm NX}^{\rm jet}$ is more significant. 
Note that the bimodal $\rho_{\rm NX}^{\rm jet}$ from EHT-only results come from the hyperparameter of MEM regularizer that smaller values tend to provide higher $\rho_{\rm NX}^{\rm jet}$. 
Likewise $\rho_{\rm NX}^{\rm jet}$, the \nxcorr can be computed on the log pixel values, $\rho_{\rm NX,log}$, after limiting the dynamic range of the images to $10^4$ (e.g., \citealt{Roelofs_2023}). 
In this way, the $\rho_{\rm NX}$ of the fainter jet features can be better determined. 
As presented in \autoref{fig:im_fiducial_nxcorr} (top, left), the $\rho_{\rm NX,log}$ also shows more significant improvement by adding eKVN to EHT. 
In addition, the residual emission is significantly reduced by the participation of the eKVN. 
For instance, the root-mean-squared (r.m.s) difference between the groundtruth and reconstructed image becomes a half for the jet reconstruction when the eKVN is included (see \autoref{fig:im_fiducial_nxcorr}, bottom). 

% Jackknife test
The EHT observing campaign carefully considers weather conditions and instrumental management to ensure optimal operations. 
Nevertheless, adverse weather conditions, particularly in certain geographic locations, can occasionally impact observing sessions, potentially reducing critical baselines. 
For instance, in the EHT2023 array, the KP and SMT in Arizona provide a unique baseline length of $\sim100$\,M$\lambda$ at 230\,GHz bridging the gap between intrasite baselines (e.g., $\sim1.6$\,M$\lambda$ by ALMA$-$APEX) and the NOEMA$-$IRAM baseline ($\sim750$\,M$\lambda$). 
Therefore, bad weather in Arizona could significantly affect observations by compromising short baseline $uv-$coverage. 
In this regard, the eKVN serves as an effective supplement with baseline lengths ranging 74$-$370\,M$\lambda$. 
To investigate the geographical impact of telescopes on image reconstruction, we implemented the jackknife test by flagging telescopes in similar regions: Chile (ALMA, APEX), \hawaii (JCMT, SMA), mainland of United States of America (USA; KP, SMT), Mexico (LMT), Greenland (GLT), and Europe (EU; NOEMA, IRAM). 
\autoref{fig:jackknife} presents the results of jackknife test with the fiducial imaging parameters. Full array images are consistent with \autoref{fig:im_fiducial_nxcorr}. 
By the eKVN participation, it is clearly found that the large scale jet can be better recovered as seen by the $\rho_{\rm NX}^{\rm jet}$ especially when telescopes in Chile, EU, or USA are missing. 

The central ring structure has also been better reconstructed in EHT$+$eKVN that is supported by not only the $\rho_{\rm NX}^{\rm ring}$ but also the ring parameters. 
To extract the ring parameters, we utilized the image domain feature extraction (IDFE) pipeline, employing the Variational Image-Domain Analysis ({\tt VIDA}; \citealt{Tiede_2022}) with the $m$F-ring model. 
This method determines the ring parameters such as radius, width, structural asymmetry (i.e., ellipticity, $\tau$), brightness asymmetry, and the position angle of the brightest spot. 
Note that the ring width is affected by the effective resolution of the image that can vary in each jackknife test, and the brightness distribution can be affected by the diffuse jet emission. 
Therefore, we compare the ring diameter and structural asymmetry to evaluate the ring structure (see \autoref{fig:idfe}). 
For reference, we present the same fitting results applied to the groundtruth image. 
In terms of the ring diameter, full array and all jackknife tests recover well the true value, $\sim39.5\,\mu$as, within $\sim5\,\mu$as uncertainties. 
This is also consistent with the measured ring diameter from the EHT observation in 2018, $43.3_{-3.1}^{+1.5}\,\mu$as \citep{eht2018_M87_P1}. 
As for the ring ellipticity, the results from EHT$+$eKVN provides more stable results than EHT-only that are close to the groundtruth value of $\tau\sim0.12$.

%%% FIGURE %%%%%%%%%%%%%%%%%%%%%%%%%%%%%%%%%%%%%%%%%%%%%%%%%%%%%%%%%%%%%%%%%%%
\begin{figure*}[ht!]
\centering 
\includegraphics[width=\linewidth]{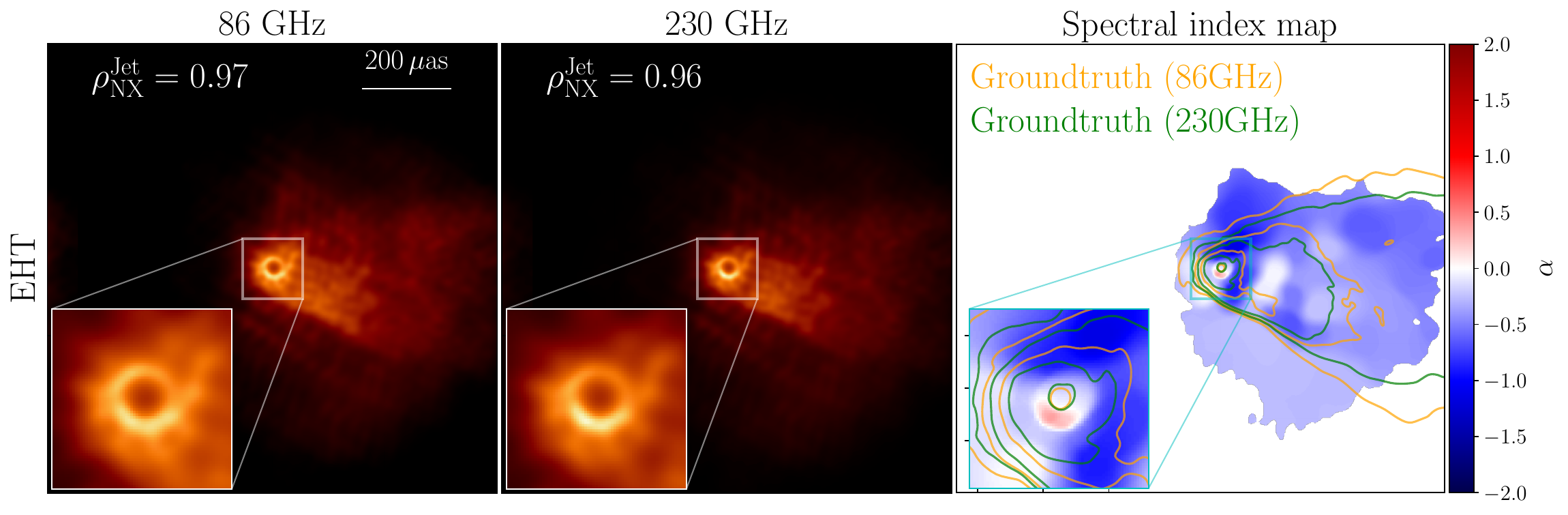}
\includegraphics[width=\linewidth]{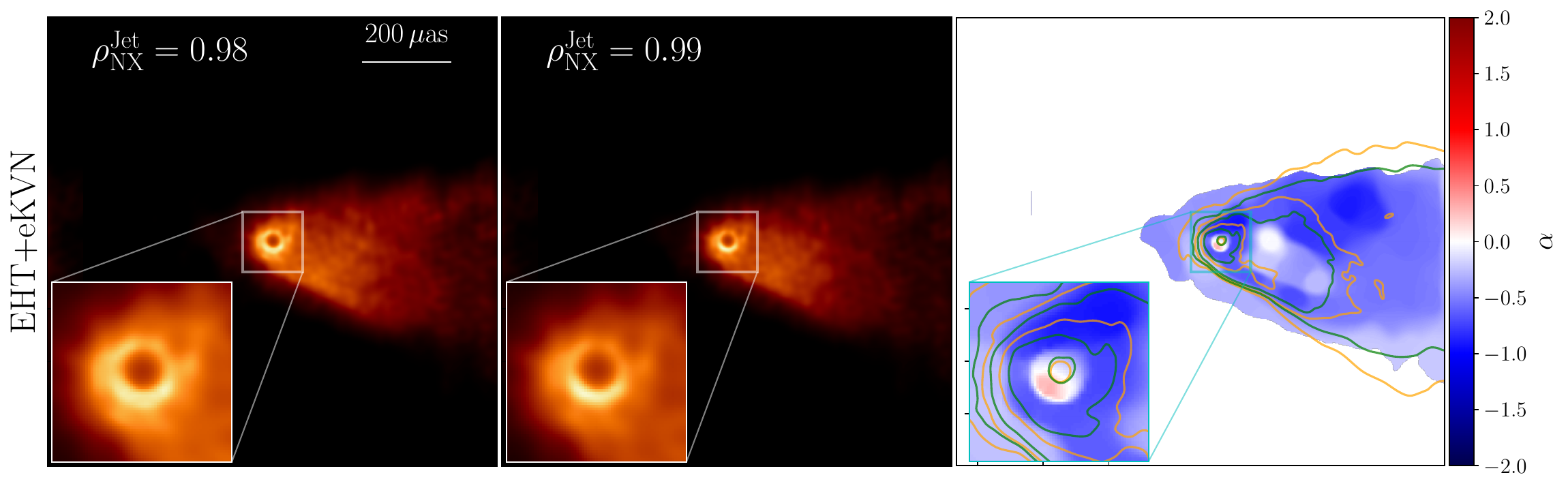}
\caption{
The multi-frequency synthesis imaging results at 86 and 230\,GHz. 
From left to right, images at 86\,GHz, 230\,GHz, and spectral index map. 
From top to bottom, reconstructed images with EHT-only and EHT$+$eKVN. 
The spectral index, $\alpha$, follows the convention of $S_{\nu}\propto\nu^{\alpha}$, where $S_{\nu}$ and $\nu$ are the flux density and observing frequency, respectively. 
The contours on the spectral index maps present the groundtruth structure at 86\,GHz (orange) and 230\,GHz (green) for comparison, after convolving with 8\,$\mu$as of cirular Gaussian beam. 
Total intensity color scale is same with \autoref{fig:im_fiducial_nxcorr}. 
}
\label{fig:mf_image}
\end{figure*}
%%% FIGURE %%%%%%%%%%%%%%%%%%%%%%%%%%%%%%%%%%%%%%%%%%%%%%%%%%%%%%%%%%%%%%%%%%%

\subsection{86$-$230\,GHz synthesis} \label{subsec:mf_synthesis}
\autoref{fig:mf_image} presents the results of multi-frequency synthesis imaging. From left to right, the images at 86\,GHz, 230\,GHz, and spectral index map are displayed respectively, from the EHT-only (top) and EHT$+$eKVN (bottom). 
The contours in the spectral index map presents the groundtruth structure at 86\,GHz (orange) and 230\,GHz (green) at a level of 0.1, 1, 10\,\% of peak flux density. 
In the images at 230\,GHz, both EHT-only and EHT$+$eKVN recover well the groundtruth structure. 
Especially from EHT-only, the jet reconstruction is much improved in terms of $\rho_{\rm NX}^{\rm Jet}$ and the reduced residual noise (see, \autoref{fig:im_fiducial_nxcorr}, for comparison). 
This is thanks to the 86\,GHz data points providing the short baselines region that play a similar role with eKVN.  
Both the central ring structure and larger scale jet are well reconstructed even at 86\,GHz, as also reported by the GMVA$+$ALMA observation \citep{lu_2023}. 
At 86\,GHz, in addition, there are no intrasite baselines (i.e., ALMA$-$APEX or JCMT$-$SMA) and the eKVN provides the shortest baseline lengths (see \autoref{fig:uvplt}). 
This implies that the eKVN can provide a better estimate of the compact flux density which is critical for reconstructing the image by discerning the dominance of the central ring compared to the fainter jet structure. 

The spectral index map was first obtained in every pixel of the image. 
Since the fluxes outside the intrinsic structure were assumed as residual noise, $\sigma_{\rm s}$, only a region where the fluxes higher than 3$\times\sigma_{\rm s}$ were used to present the spectral index map. 
Note that $\sigma_{\rm s}$ was estimated outside the source structure that was guided by the groundtruth image. 
As a result, the southern part of a ring structure is optically thick and the rest of features are optically thin, which is well consistent with the groundtruth (see, \autoref{fig:groundtruth}) in both EHT-only and EHT$+$eKVN results. 
As shown in \autoref{fig:mf_image}, however, the threshold indicates that the high fidelity region in the spectral index map is more reliable in the EHT$+$eKVN result than that in the EHT-only. 
This demonstrates well the importance of the short baselines that can be filled with the eKVN to avoid potential biases in the reconstruction of larger scale structure. 
As shown in the images at each 86 and 230\,GHz, as well as the spectral index map, the EHT-only result provides more diffuse emissions outside the real structure that reduce the dynamic range of images. 
The artificial noise features are clearly removed by the addition of eKVN so that the intrinsic structure at larger scale are better constrained. 
This will be even more critical for the fainter sources or any targets with potential jet structure to be discovered (e.g., \sgra; \citealt{moscibrodzka_2014, brinkerink_2015}).

%%%%%%%%%%%%%%%%%%%%%%%%%%%%%%%%%%%%%%%%%%
\section{Discussions and Conclusions} \label{sec:summary}

% Summary 
In this study, we made quantitative assessments of the improvement in EHT performance by incorporating the eKVN. 
Since the first image of the black hole shadow in \meight by the EHT, it has been aimed to reconstruct the jet alongside the black hole shadow. 
As introduced in \autoref{sec:intro}, understanding the jet launching mechanism -- whether black hole driven or accretion flow driven -- is of great importance and represents a longstanding question. 
% EHT + KVN
The participation of GLT in the 2018 observations improved the image quality and helped to reveal the variations in the brightness distribution on the ring, but it was still insufficient to recover the diffuse jet emission \citep{eht2018_M87_P1}. 
Later EHT observations with improved $uv-$coverage by the participation of KP and NOEMA may have a potential to achieve this but it remains uncertain due to the lack of short baseline lengths which are crucial for recovering larger scale structures. 
In this regard, the eKVN can play a unique and important role as part of the EHT providing short baseline lengths corresponding to angular scale of up to $\sim2\,$mas, as we verified with the synthetic data based on the simulated \meight model including jet. 
The improved $uv-$coverage of EHT compared to 2017 and 2018 is already recovering the central ring and jet structure well. Nevertheless, when the eKVN is included in the EHT observations, (i) the residual noise is reduced to almost half, (ii) it is less affected by missing telescopes, and (iii) the spectral index distribution is better constrained, particularly along the large-scale jet. 
While there are slight improvements in the ring structure, the major advantages provided by the eKVN are related to jet reconstruction, thanks to the provision of six more baselines shorter than 1\,G$\lambda$ which has been two in EHT-only observations (KP$-$SMT and NOEMA$-$IRAM). 

% Other arrays  
The 230\,GHz receiver has already been equipped at KYS, participating in the EHT observational campaign in April, 2024. 
Currently it is alternated with the 129\,GHz receiver, and 120$-$230\,GHz dual-band receiver is planned for the future. 
The KPC will be the next to be equipped with the 230\,GHz receiver in 2024$-$2025. 
As for KUS and KTN, the timeline is still uncertain but hopefully being completed in a few years. 
Once the 230\,GHz receiver is installed at all eKVN telescopes, the full eKVN is expected to make a significant contribution in various next-generation VLBI arrays including the EHT and ngEHT. 
For instance, with the East Asian VLBI Network (EAVN; e.g., \citealt{Akiyama_2022}), the EAVN-hi \citep{Asada_2017} has been planned with GLT, JCMT, and Seoul Radio Astronomical Observatory (SRAO; \citealt{Shin_2020}) operating at 86 and 230\, GHz, enabling dense monitoring observations. 
In addition, the EHT \meight movie campaign has been planned to find the time variation of the ring feature \citep{Johnson_2023, eht2018_M87_P1} together with the innermost jet structure. 
Their relative motions will provide hints of the black hole spin and possible jet precession \citep{Cui_2023}, as well as insights into the connection between the black hole and the jet through simultaneous imaging \citep[e.g.,][]{lu_2023}. 
For consistent reconstruction of a jet throughout the monitoring observations, securing enough numbers of short baselines is important that can be provided by the eKVN. 
As for the ngEHT, Owens Valley Radio Observatory (OVRO) 40\,m telescope and Haystack Observatory 37\,m telescope in USA are planning to join at both 86 and 230\,GHz. 
The reference array of ngEHT additionally includes more than ten telescopes, such as 
AMT (Gamsberg, Namibia), BOL (La Paz, Bolivia), CNI (La Palma, Canary Islands), JELM (Wyoming, USA), KILI (Kilimanjaro, Tanzania), LCO (Coquimbo, Chile), LLA (Salta, Argentina), SGO (Santiago, Chile), SPM (Baja California, Mexico), and SPX (Fieschertal, Switzerland) \citep{Doeleman_2023}. 
In line with this, the eKVN continues to utilize the unique geographical locations in East Asia that are compatible with the reference array. 
In addition, the (ng)EHT$+$eKVN will provide a better chances to explore new sites in Asia (possibly, e.g., China, India, Thailand). This will have long-term impact on the global VLBI array. 

% Further science cases 
% Sgr A*
When it comes to \sgra, overcoming the fast structural variation on a timescale of tens of minutes is now required, which can be achieved by movie reconstruction using dynamic imaging techniques \citep{Bouman_2017, johnson_2017_dynam, SgraP3}. 
For this purpose, it is important to have sufficient $uv-$coverage in a short time range (e.g., snapshot; see \ref{app:sgra} for more details). 
Therefore, the EHT$+$eKVN will also improve the movie fidelity in dynamic imaging for \sgra which will reveal fast motions along the black hole shadow for the first time. 
With the future planned upgrades, more detailed studies toward SMBH and AGN jet will also be available. 
For instance, detection of fainter horizon-scale targets through phase referencing observations and astrometry of bright targets through SFPR. 
With the polarimetric observations, Faraday rotation map in faint jet can be measured. 
Lastly, possible science cases with eKVN-only observations will also be explored in a forthcoming paper. 

%%% ACKNOWLEDGMENTS (IF ANY) %%%%%%%%%%%%%%%%%%%%%%%%%%%%%%%%%%%%%%%%

\acknowledgments

%We are grateful to anonymous referee for constructive comments and suggestions that were helpful to improve the manuscript. 
% Funding
I.C. is supported by the KASI-Yonsei Postdoctoral Fellowship. 
This work is supported by the National Research Foundation of Korea (NRF) grant funded by the Korean government (Ministry of Science and ICT) for J.P. (RS-2024-00449206), J.Y.K. (2022R1C1C1005255, and RS-2022-00197685), S.T. (2022R1F1A1075115), and A.Chung (2022R1A2C100298213, and RS-2022-NR070872). 
J.P. also acknowledges support by the POSCO Science Fellowship of POSCO TJ Park Foundation. 
A.Chael is supported by the Princeton Gravity Initiative. 
S.I. is supported by Hubble Fellowship grant HST-HF2-51482.001-A awarded by the Space Telescope Science Institute, which is operated by the Association of Universities for Research in Astronomy, Inc., for NASA, under contract NAS5-26555. 
%A.Chung acknowledge support by the NRF, grant Nos. \icedt{2022R1A2C100298213, and RS-2022-NR070872}. 
% Softwares
For synthetic data generation, the \ngehtsim software \citep{Pesce_2024, Pesce_2024zndo} is used. 
For imaging and analysis, the \ehtim software \citep{Chael_2018, Chael_2023} is used.

%%% APPENDICES (IF ANY) %%%%%%%%%%%%%%%%%%%%%%%%%%%%%%%%%%%%%%%%%%%%%

\appendix
\section{eKVN sites and their partial participation in the EHT}
\label{app:ekvn_sites}

\begin{figure*}[t]
\includegraphics[width=0.9\linewidth]{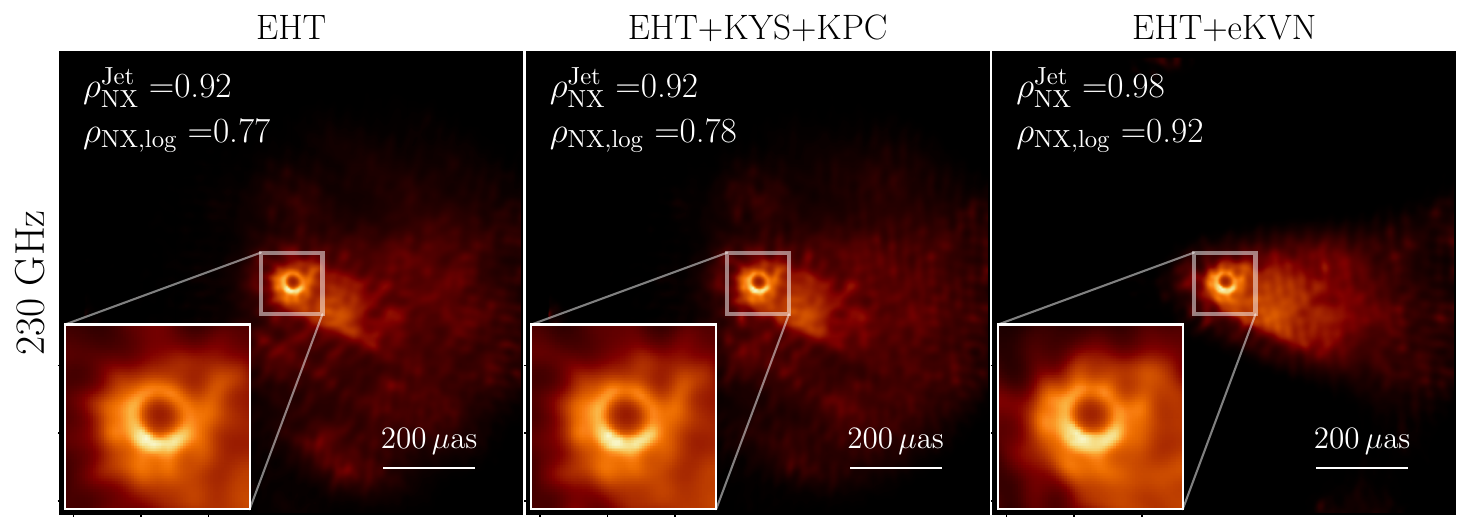}
\caption{Same as \autoref{fig:im_fiducial_nxcorr} (top) but with the EHT$+$KYS$+$KPC.}
\label{fig:single_2kvn}
\end{figure*}

\begin{figure*}[t]
\centering 
\includegraphics[width=\linewidth]{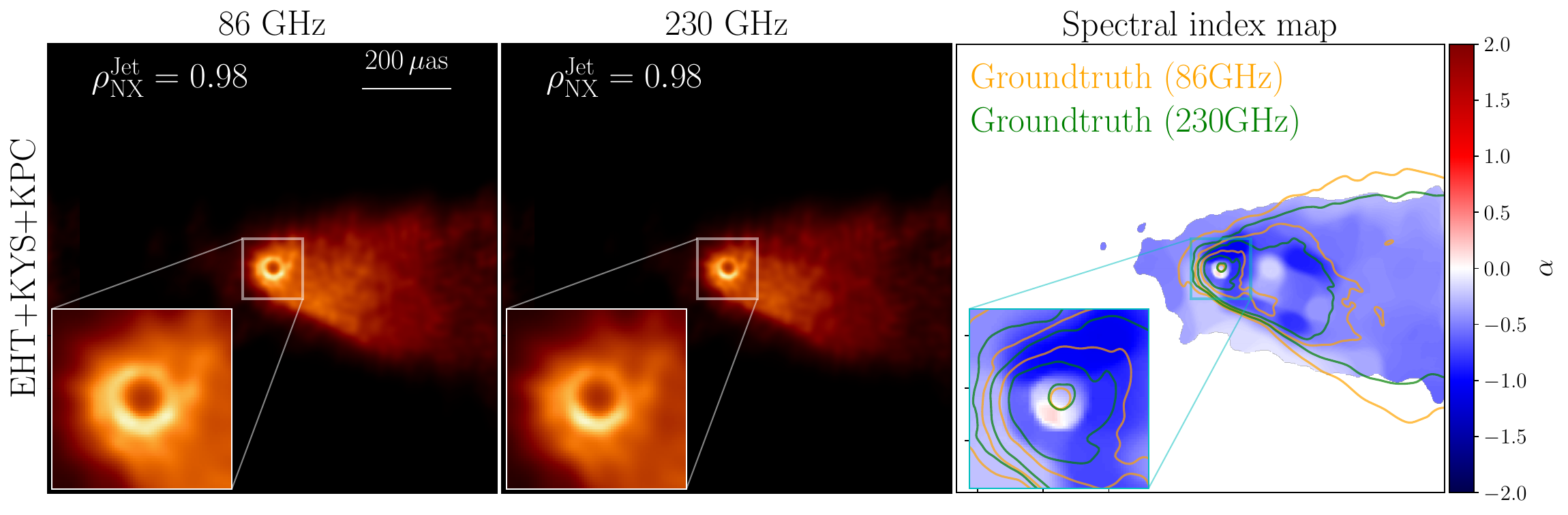}
\caption{Same as \autoref{fig:mf_image} but with the EHT$+$KYS$+$KPC.}
\label{fig:mf_image_2kvn}
\end{figure*}

\begin{figure}[ht!]
\centering 
\includegraphics[width=\linewidth]{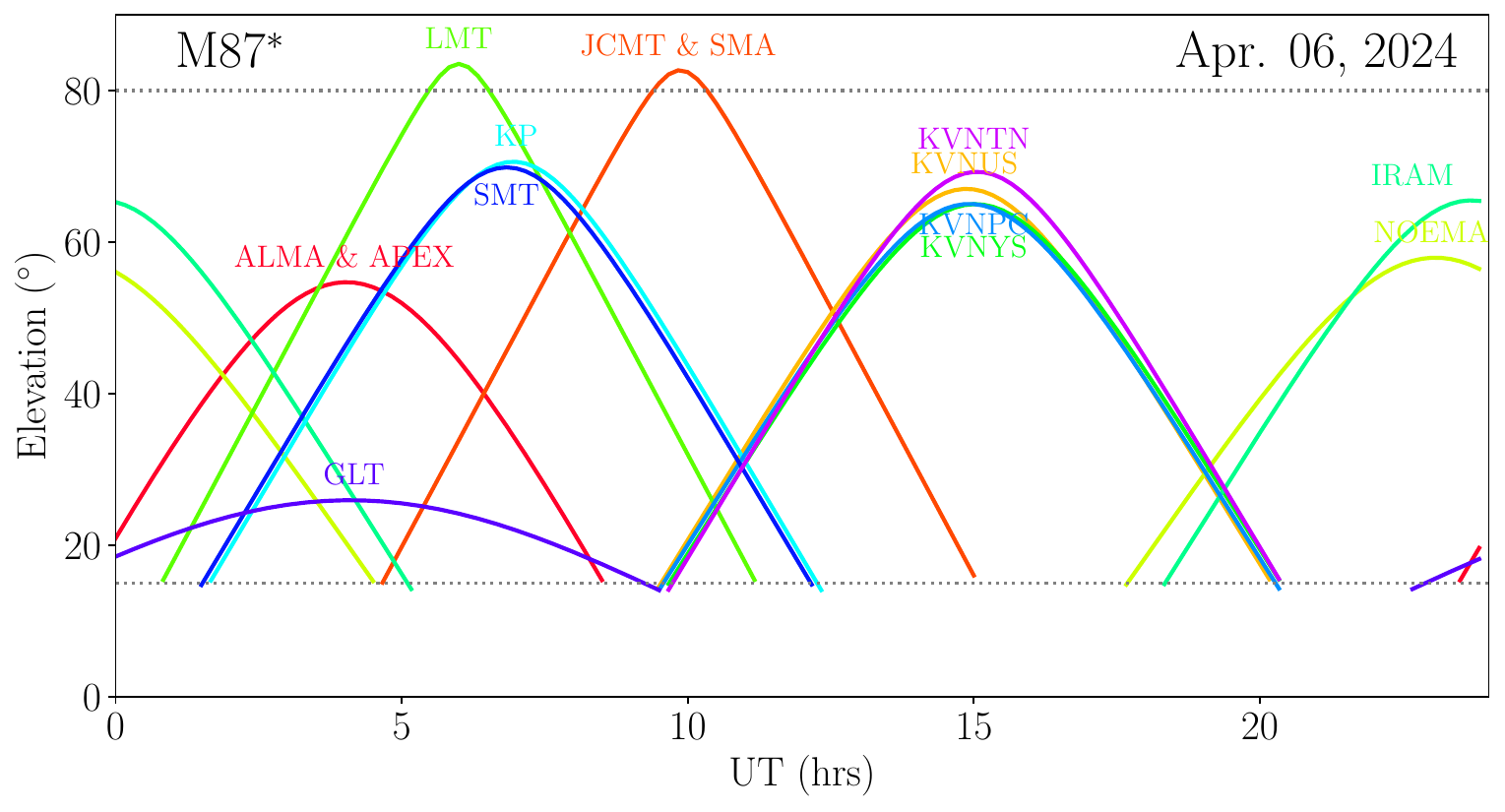}
\caption{
Elevation of \meight along with observing time at each telescope on April~6, 2024. 
The synthetic data for this study is generated based on this observation. 
Horizontal, dotted lines present the elevation limits of 15$^\circ$ and 80$^\circ$.
}
\label{fig:elev_m87}
\end{figure}

\begin{figure*}[t]
\centering 
\includegraphics[width=\linewidth]{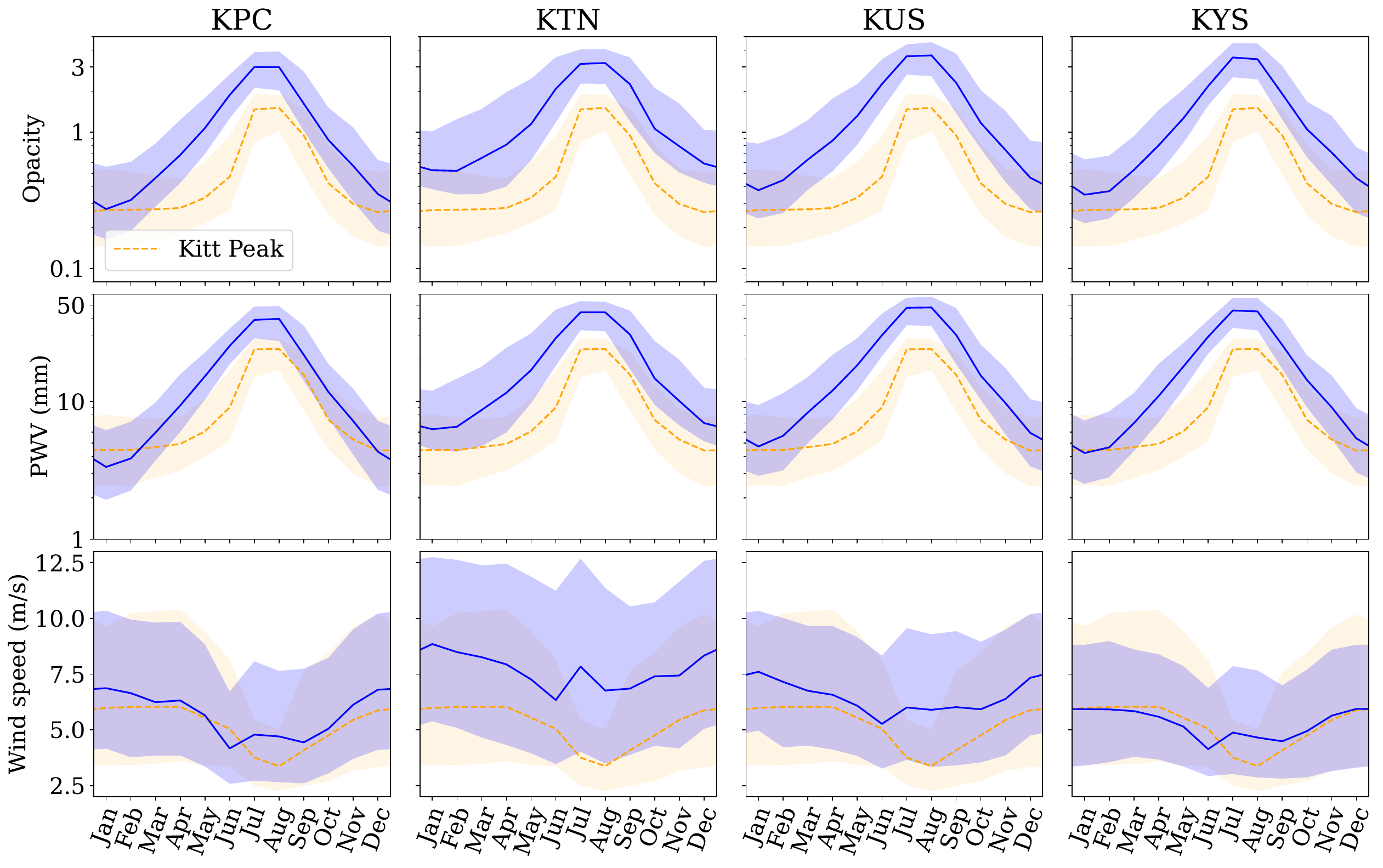} 
\caption{
Generic weather condition over a year at each eKVN telescope site from the \ngehtsim database: (from top to bottom) atmospheric opacity to the zenith at 230\,GHz, precipitable water vapor (mm), and wind speed (m/s). 
Blue, solid line is mean estimate and shaded area represents $1\sigma$ range. 
Same values at Kitt Peak (KP) are shown together in orange color for comparison. 
}
\label{fig:weather}
\end{figure*}

Three of the KVN telescopes have been successfully operating at frequencies up to 130\,GHz, and the KPC has been newly built in 2023 that has CTR at 22, 43, and 86\,GHz. 
In addition, as introduced in this study, the 230\,GHz receiver has already been installed at KYS and will be available at KPC starting in 2025. 
For the other two telescopes, KUS and KTN, the 230\,GHz receiver may be installed in the coming years.
Based on this timeline, we present the imaging simulation results when only KYS and KPC participate in the EHT at 230\,GHz. Since only one short baseline is added, the improvement in the image is limited (\autoref{fig:single_2kvn}). 
However, this changes dramatically in the multi frequency synthesis imaging because all four eKVN telescopes can operate at 86\,GHz even if only two sites are active at 230\,GHz. 
As a result, the images are almost identical to those produced with full eKVN participation (see, \autoref{fig:mf_image_2kvn}). 
This highlights that the current contribution of two KVN telescopes to the EHT can already enhance imaging performance when 86\,GHz is simultaneously observed. 
In addition, it shows the importance of future participation of full eKVN in the EHT and ngEHT, given its unique geographical locations in East Asia (see, \autoref{tab:kvnspec} and \autoref{fig:elev_m87}). 
Note that the exact antenna positions of eKVN are routinely monitored by GPS and geodetic VLBI observations in collaboration with VLBI Exploration of Radio Astrometry (VERA) in Japan. 
While the EHT and ngEHT plan to observe up to 345\,GHz, the eKVN cannot operate at this frequency due to limited weather condition but it is still doable at 230\,GHz. 
\autoref{fig:weather} presents the seasonal weather condition at each of eKVN telescope site including atmospheric opacity to the zenith at 230\,GHz, precipitable water vapor in mm, and wind speed in m/s. For comparison, the same values at Kitt Peak (KP) are shown together which are comparable to eKVN sites. 

%%% FIGURE %%%%%%%%%%%%%%%%%%%%%%%%%%%%%%%%%%%%%%%%%%%%%%%%%%%%%%%%%%%%%%%%%%%
%%% FIGURE %%%%%%%%%%%%%%%%%%%%%%%%%%%%%%%%%%%%%%%%%%%%%%%%%%%%%%%%%%%%%%%%%%%

%%% FIGURE %%%%%%%%%%%%%%%%%%%%%%%%%%%%%%%%%%%%%%%%%%%%%%%%%%%%%%%%%%%%%%%%%%%
%%% FIGURE %%%%%%%%%%%%%%%%%%%%%%%%%%%%%%%%%%%%%%%%%%%%%%%%%%%%%%%%%%%%%%%%%%%

%%%%%%%%%%%%%%%%%%%%%%%%%%%%%%%%%%%%%%%%%%%%%%%%%%%%%%%%%%%%%%%%%%%%%%%%%%%%%%%%%%%%%%%%%%%%%%
\begin{table}[ht!]
\centering
\caption{Coordinates of eKVN telescopes}
\begin{tabular}{ccccc}
\hline
\hline
Telescope & Longitude & Latitude & Altitude \\
 & ($^{\circ}$ $'$ $''$) & ($^{\circ}$ $'$ $''$) & (m) \\
\hline
KYS & 126:56:27.4 & 37:33:54.9 & 139  \\
KUS & 129:14:59.3 & 35:32:44.2 & 170 \\
KTN & 126:27:34.4 & 33:17:20.9 & 452 \\
KPC & 128:26:55.1 & 37:32:00.1 & 557 \\
\hline 
\end{tabular}
\label{tab:kvnspec}
\end{table}
%%%%%%%%%%%%%%%%%%%%%%%%%%%%%%%%%%%%%%%%%%%%%%%%%%%%%%%%%%%%%%%%%%%%%%%%%%%%%%%%%%%%%%%%%%%%%%

%%% FIGURE %%%%%%%%%%%%%%%%%%%%%%%%%%%%%%%%%%%%%%%%%%%%%%%%%%%%%%%%%%%%%%%%%%%
%%% FIGURE %%%%%%%%%%%%%%%%%%%%%%%%%%%%%%%%%%%%%%%%%%%%%%%%%%%%%%%%%%%%%%%%%%%

%%% FIGURE %%%%%%%%%%%%%%%%%%%%%%%%%%%%%%%%%%%%%%%%%%%%%%%%%%%%%%%%%%%%%%%%%%%
%%% FIGURE %%%%%%%%%%%%%%%%%%%%%%%%%%%%%%%%%%%%%%%%%%%%%%%%%%%%%%%%%%%%%%%%%%%

\section{Effective angular resolution}
\label{app:beam} 

\begin{figure*}[t]
\includegraphics[width=\linewidth]{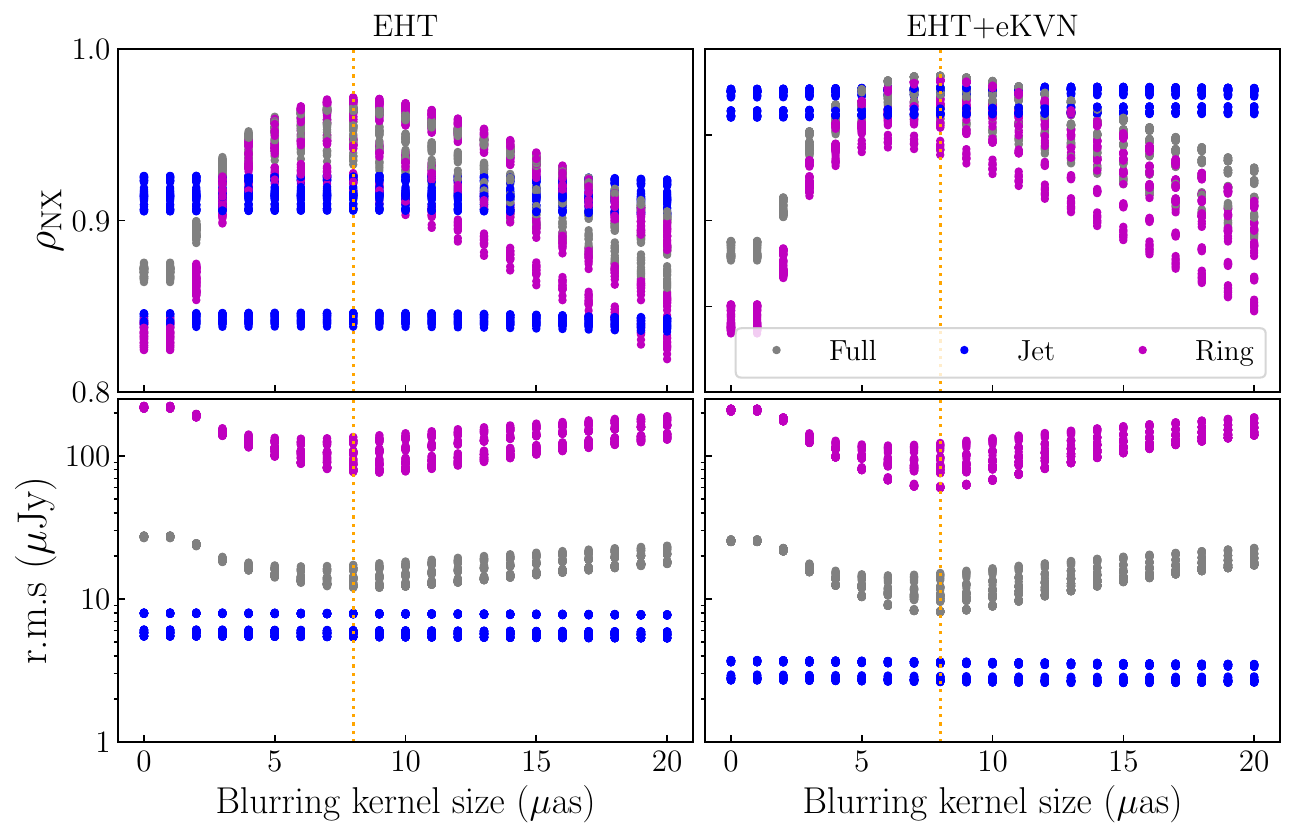} 
\caption{
(Top) The \nxcorr between the reconstructed images and a groundtruth image, as a function of the convolution Gaussian kernel size to latter, for EHT-only (left) and EHT$+$eKVN (right) results. 
(Bottom) The r.m.s of the reconstructed images. 
The color-code is same as \autoref{fig:im_fiducial_nxcorr} $-$ \nxcorr of full image (gray), jet (blue), and ring (magenta) structures. 
The vertical-dotted, orange line presents the effective resolution of reconstructed images that provides the highest \nxcorr and lowest r.m.s. 
}
\label{fig:nxcorr_beam}
\end{figure*}

The nominal resolution of reconstructed images is constrained by the diffraction limit, typically determined by the longest baseline length. 
For both the EHT and EHT$+$eKVN arrays in our study, this limit corresponds to $\sim25\,\mu$as at 230\,GHz. 
By applying the forward modeling such as the RML imaging, however, it is possible to achieve a better effective angular resolution, known as super-resolution. 
This can be achieved to factor of $1/3$ to $1/2$ the nominal resolution when the S/N of visibility data is sufficiently high \citep[e.g.,][]{chael_2016}. 
To practically determine the effective resolution of the reconstructed images, we estimated the \nxcorr between the images and a groundtruth image, by convolving latter with a circular Gaussian kernel of varying sizes up to 20\,$\mu$as (see, \autoref{fig:nxcorr_beam}). 
By this way, we found that a beam size of 8\,$\mu$as yields the highest \nxcorr corresponding to $\sim1/3$ of the nominal resolution in our data. 
The \nxcorr comparison in the main context are then implemented after convolving the groundtruth image with the effective angular resolution. 

%%% FIGURE %%%%%%%%%%%%%%%%%%%%%%%%%%%%%%%%%%%%%%%%%%%%%%%%%%%%%%%%%%%%%%%%%%%
%%% FIGURE %%%%%%%%%%%%%%%%%%%%%%%%%%%%%%%%%%%%%%%%%%%%%%%%%%%%%%%%%%%%%%%%%%%

\section{EHT$+$eKVN towards \sgra}
\label{app:sgra}

\begin{figure*}[ht!]
\centering 
\includegraphics[width=\columnwidth]{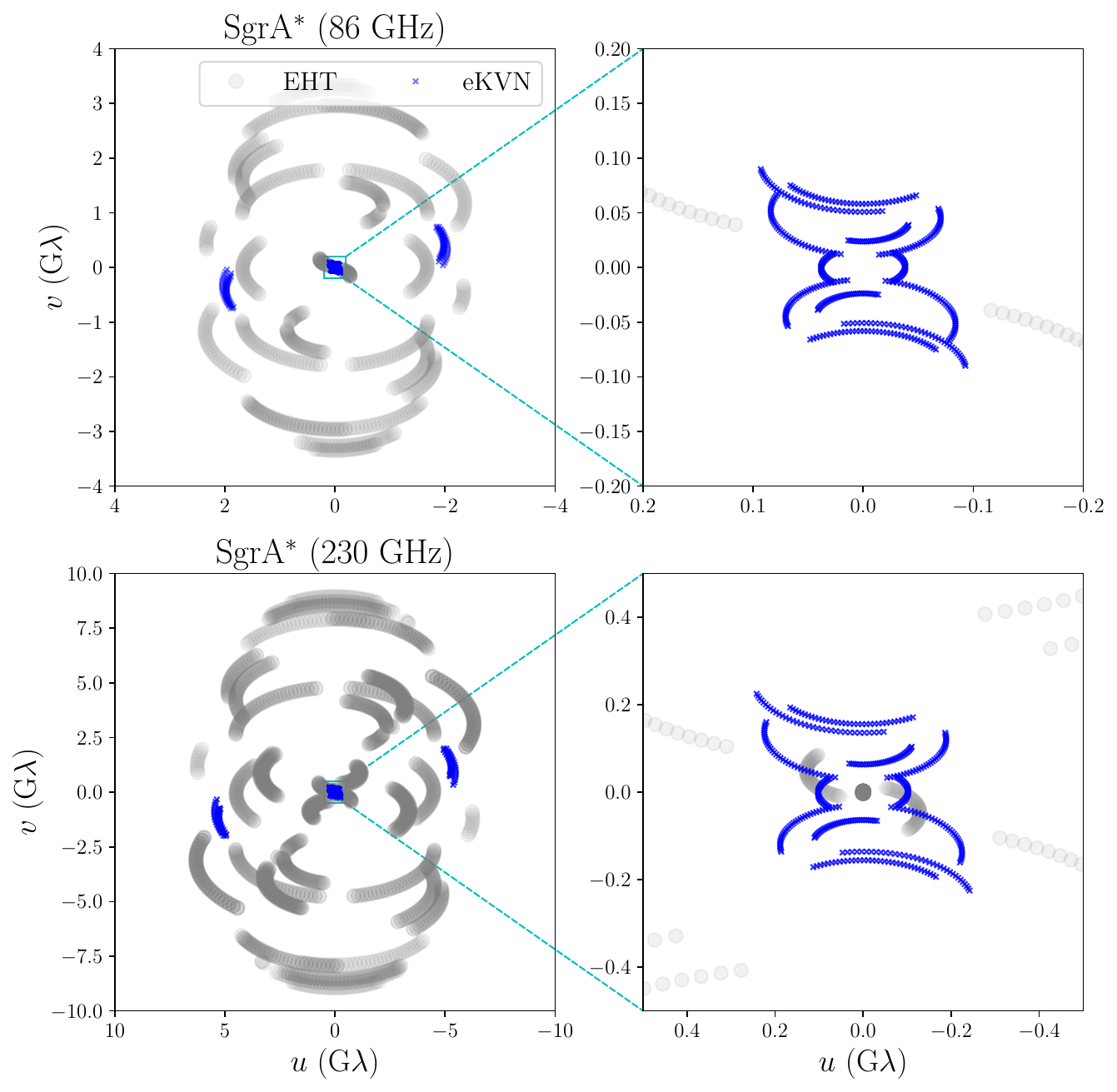}
\includegraphics[width=\columnwidth]{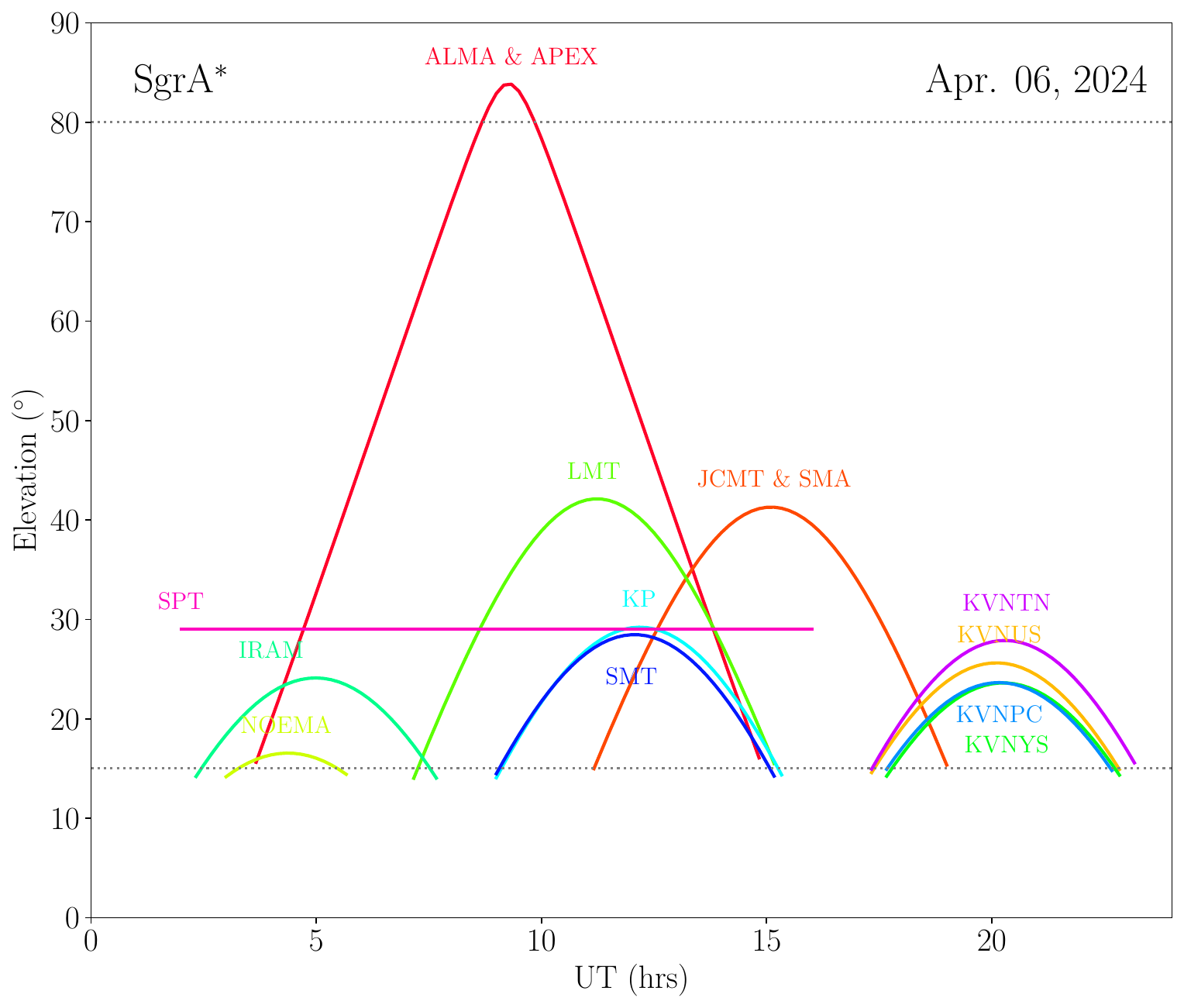}
\caption{
(Left) The $uv-$coverage of \sgra at 86\,GHz (top) and 230\,GHz (bottom). Left and right panels show the full and short baselines range corresponding to the eKVN baselines, respectively. 
(Right) The elevation of \sgra along with observing time on April~6, 2024. 
}
\label{fig:sgra_uv_elev}
\end{figure*}

\sgra is the closest SMBH from Earth that extends the largest angular diameter of the black hole shadow, $51.8\,\pm\,2.3\,\mu$as \citep{SgraP1}, and the size of intrinsic structure is wavelength dependent at mm to cm \citep[e.g.,][]{Cho_2022}. 
The black hole mass, $\sim4\times10^6\,M_\odot$, is about 1,000 times less than \meight which results in much shorter light crossing time of a few to tens of minutes, while it takes days for \meight. 
From several hours of VLBI observations, therefore, the image we obtain corresponds to time averaged one while its structure is varying. 
To overcome this, as described in \autoref{sec:summary}, dynamic imaging has been developed and applied to the EHT data (\citealt{SgraP3}; EHT~Collaboration~in~prep.). 
The performance of dynamic imaging relies on the $uv-$coverage for short time interval (i.e., snapshot) so it is important to have more telescopes across the observing time. 
\autoref{fig:sgra_uv_elev} presents the $uv-$coverage toward \sgra from EHT$+$eKVN on the same day with the synthetic data for \meight. 
The elevation along with observing time is shown together. 
Currently there are limited common visibilities between EHT and eKVN but more telescopes will fill the gaps as a part of ngEHT. 
In addition, eKVN provides the shortest baselines at 86\,GHz and effectively fills the missing short baseline ranges at 230\,GHz. 
This is particularly important to constrain the varying flux density of \sgra, for instance having been provided by ALMA light curve \citep{Wielgus_2022}. 

%%% FIGURE %%%%%%%%%%%%%%%%%%%%%%%%%%%%%%%%%%%%%%%%%%%%%%%%%%%%%%%%%%%%%%%%%%%
%%% FIGURE %%%%%%%%%%%%%%%%%%%%%%%%%%%%%%%%%%%%%%%%%%%%%%%%%%%%%%%%%%%%%%%%%%%

%%% CALL LIST OF REFERENCES (natbib STYLE) %%%%%%%%%%%%%%%%%%%%%%%%%%
% \bibliography{jkas-sample}

%\begin{thebibliography}{}
\bibliography{paper_jkas.bib}

%%% PUT YOUR REFERENCES HERE %%%%%%%%%%%%%%%%%%%%%%%%%%%%%%%%%%%%%%%%

%\bibitem[Salpeter(1955)]{salpeter1955} Salpeter, E. E. 1955, The Luminosity Function and Stellar Evolution, ApJ, 121, 161

%%% END LIST OF REFERENCES %%%%%%%%%%%%%%%%%%%%%%%%%%%%%%%%%%%%%%%%%%

%\end{thebibliography}

\end{document}